\documentclass[twocolumn,trackchanges]{aastex701}

\newcommand{\tighttile}{}

\begin{document}

\title{A Search for Transit Duration Variations in M dwarf Multi-Planet Systems}

\author{Kohhei Bessho}
\affiliation{Department of Physics, Waseda University, 3-4-1 Ookubo, Shinjuku-ku, Tokyo, 1698855, Japan}
\email[show]{koheit@akane.waseda.jp}

\author[0000-0002-3247-5081]{Sarah Ballard}
\affiliation{Department of Astronomy, University of Florida, Gainesville, FL 32611}
\email{sarahballard@ufl.edu}

\author{Natalia Guerrero}
\affiliation{Department of Astronomy, University of Florida, Gainesville, FL 32611}
\email{sarahballard@ufl.edu}


\begin{abstract}
The nominal habitable zone for exoplanets orbiting M dwarfs lies close to the host star, making dynamical considerations especially important. One consequence of this proximity is the expectation of spin synchronization, with implications for atmospheric circulation. Several mechanisms can maintain non-zero obliquities over long timescales in compact multiplanet systems, including capture into Cassini State 2 (CS2) and other forms of secular spin–orbit coupling; such pathways are plausible in the orbital architectures of close-in M-dwarf planets. In this study, we search for transit duration variations (TDVs) consistent with the nodal precession rates predicted by Laplace–Lagrange secular theory in compact M-dwarf multiplanet systems. Our sample includes 23 exoplanets orbiting 12 stars. We compare recent, high-precision transit durations obtained from JWST white-light curves with measurements published at the discovery epoch and afterward. The resulting transit duration variation ranges from seconds to minutes, and we fit a linear trend to duration versus time for each planet. All systems are consistent with flat (no TDV) at the 3$\sigma$ level. The strongest candidate is TRAPPIST-1d, whose fitted slope is discrepant from zero with 2.2$\sigma$ confidence. We calculate the expected TDV signals predicted by secular precession and compare them to the observed limits. Our null detection is consistent with the low–impact-parameter regime, where theoretical TDVs are only a few seconds per decade and below our sensitivity. Higher–impact-parameter configurations predict substantially larger TDVs and are disfavored: under uniformly distributed geometries, at least half of the allowed configurations would be excluded. 
\end{abstract}

\section{Introduction}

The atmospheres of small planets orbiting M dwarfs are rapidly moving into observational focus. The ubiquity of planets orbiting M dwarfs, coupled with the favorable detectability of planetary signals, make them compelling sites for investigating habitability \citep{tarter_reappraisal_2007, dressing_occurrence_2013, mulders_increase_2015, shields_habitability_2016, Hardegree19}. 
With the advent of JWST, atmospheric characterization has already yielded striking results, including the potential detection of dimethyl sulfide in K2-18 b \citep{madhusudhan_new_2025}, as well as detections of carbon-bearing molecules and cloud signatures in other sub-Neptune systems (e.g., K2-18, GJ 1214, TRAPPIST-1, \citealt{madhusudhan_new_2025,gao_hazy_2023,schlawin_possible_2024,zieba_no_2023}). 
Such measurements underscore both the opportunity and the challenge: to assess whether M-dwarf planets retain their atmospheres over gigayear timescales, and to determine whether those atmospheres are synchronous and tidally locked to their host stars.

M dwarfs furnish very different conditions for their planetary systems than Sunlike stars, in a way that makes dynamical information very important. Their lower luminosity implies a much more compact habitable zone \citep{kopparapu_habitable_2013, kopparapu_habitable_2014} in which tidal locking and circularization are commonly assumed \citep{barnes_tidal_2017, kopparapu_inner_2016, hu_role_2014, yang_strong_2014}.  The tidal heating associated with close-in exoplanets, whether due to eccentricity or obliquity, is relevant in the orbital separation range associated with potential habitability for M dwarfs \citep{barnes_tidal_2013, Colose21, driscoll_tidal_2015, valente_tidal_2022, guerrero_plausibility_2024}. Basic long-term atmospheric stability, let alone the detailed geological and atmospheric implications in these configurations, poses major challenges for understanding M dwarf exoplanets as sites for life \citep{Joshi97, heng_gliese_2011,  edson_carbonate-silicate_2012, yang_strong_2014, Yang14}. 

However, recent studies have challenged the default assumption of tidal locking for some close-in exoplanets. Potential mechanisms for resisting full synchronous rotation include thermal tides \citep{Leconte15} and dynamical interactions with neighbors \citep{vinson_spin_2017}. Tidal locking need not be a default assumption if the planet retains non-zero spin obliquity \citep{Millholland19_obliquity, valente_tidal_2022}. Specifically, this non-zero spin obliquity can persist over long periods in the case of some Cassini State (CS) resonances, specifically CS2 \citep{winn_obliquity_2005, Fabrycky07a, Millholland19_obliquity, Su2020, su_dynamics_2022}. This specific resonance of spin precession and orbital precession may result from multiple pathways, including capture during early migration (which can drive obliquities high) or in scenarios after primordial impacts induce high obliquities, which persist after spin-down \citep{valente_tidal_2022, guerrero_plausibility_2024}.  Establishing whether small planets orbiting M dwarfs reside in Cassini States, via constraints on their orbital precession, may be a precious clue to their rotation. In this context, Laplace–Lagrange secular theory provides the expected nodal precession rates in compact multiplanet systems \citep{murray_solar_1999, pu_eccentricities_2018}, offering the timescales against which the feasibility of such long-lived obliquity states can be evaluated.

The hypothesized orbital precession can be measured, in theory, via changes to the duration of transit (hereafter $\tau$; see e.g. \citealt{Miralda02}). Most of the confirmed TDV signals in \textit{Kepler} data are among large planets \citep{holczer2016transit, Millholland21}, whose durations identifiably changed over the 4-year mission baseline. But they may be potentially common and also newly detectable in systems of multiple short-period planets. In the CS2 scenario, the rate of change of the longitude of the ascending node, $\dot{\Omega}$, is equivalent to the Laplace–Lagrange nodal recession frequency $g$, itself determined by the spacing between planets and their neighbors, as well as their masses \citep{Millholland19_obliquity}. Studies of the rate of $g$ for multi-planet systems put the value on the order of 10$^{-4}$--10$^{-2}$ rad yr$^{-1}$ \citep{Millholland19_obliquity, guerrero_plausibility_2024}, roughly 5$\times10^{-3}$--$5\times10^{-1}$ degrees/year. Tens of percents of these systems are predicted to have nodal eigenfrequencies high enough that TDVs of order $\sim$10 minutes per decade are expected \citep{boley_transit_2020}. As in \cite{Millholland21} and other studies of TDVs over $\sim$years timescales, we note that we are sensitive only to an approximately linear drift in transit duration over a small snapspot of the full sinusoid. 

While \textit{Kepler} stars are largely too dim for detailed follow-up, many of the M dwarf multi-planet systems detected by \textit{K2} and subsequently \textit{TESS} were identified 5-10 years ago-- transit duration uncertainty on the order of minutes is typical \citep{sagear_orbital_2025}. Over a $\sim$10 yr baseline from discovery epoch to today, a change such a change is potentially detectable via high-precision transit duration measurements. The James Webb Space Telescope \citep{gardner_james_2006} offers a key opportunity here, as M dwarfs in multi-planet systems are already sites of active atmospheric study-- often for multiple planets in the same system. The exquisite precision of the white-light curves from recent JWST observations allows for a newly tractable experiment to measure $\Delta\tau$ in a large sample of M dwarf planets. 

This manuscript is organized as follows. In Section \ref{sec:methods}, we describe the identification of our sample of M dwarf multi-planet systems. We go on to detail the procedure of extracting the white-light curves from JWST NIRSpec data and measuring the resulting transit durations. In Section \ref{sec:analysis}, we calculate the change between the JWST transit durations and previous duration measurements. We also consider the results in light of predictions for the nodal precession rate. We discuss our findings in Section \ref{sec:discussion}, contextualizing a null result for TDVs in this sample in light of the precision of the light curves and the baseline of observations. We summarize and conclude in Section \ref{sec:conclusion}. 

\section{Methods} \label{sec:methods}

\subsection{Planet Sample Selection} \label{planetsampleselection}

We constructed our sample for this study by identifying (1) planets in M-dwarf multi-planet systems, where at least one planet is transiting, and (2) the public availability of JWST NIRSpec \citep{jakobsen_near-infrared_2022} Bright Object Time Series (BOTS, \citealt{birkmann_near-infrared_2022}) observations of those planets in the MAST archive. NIRSpec observations employ a large number of possible configurations across wavelength range and spectral resolution. With 7 different dispersive elements available, these configurations span wavelength ranges from 0.6--5.3\,$\mu$m. The full sample included in this study is summarized in Table \ref{table:planetsample}. All observations included in our sample were obtained using the S1600A1 aperture and one of the PRISM/CLEAR, G395H/F290LP, or G235H/F170LP disperser–filter combinations; these combinations are indicated in the Table. In total, we have included 12 multi-planet systems around M dwarfs, comprising 23 transiting planets with suitable JWST data products. The baseline between discovery epoch and the date of JWST observations spans 3--10 years for our sample, with an average of 5 years. 

\begin{deluxetable*}{ccccccccc}[ht!]
\tablecaption{JWST Observations Included in Sample\label{table:planetsample}}
\tablehead{
    \colhead{Planet Name} & 
    \colhead{Discovery Observation Span} & 
    \colhead{Observation Year} & 
    \colhead{Instrument} & 
    \colhead{Exp Type} & 
    \colhead{Optical Element} & 
    \colhead{Pro Obs \#}
}
\startdata
Trappist-1 b & 2015 \citep{2016Natur.533..221G} & 2024 & NIRSpec & BOTS & CLEAR;PRISM & 2420 5 \\ 
Trappist-1 c & 2015 \citep{2016Natur.533..221G} & 2024 & NIRSpec & BOTS & CLEAR;PRISM & 2420 5 \\ 
Trappist-1 d & 2016 \citep{2017Natur.542..456G} & 2022 & NIRSpec & BOTS & CLEAR;PRISM & 1201 111 \\ 
Trappist-1 e & 2016 \citep{2017Natur.542..456G} & 2023 & NIRSpec & BOTS & CLEAR;PRISM & 1331 2 \\ 
Trappist-1 g & 2016 \citep{2017Natur.542..456G} & 2022 & NIRSpec & BOTS & CLEAR;PRISM & 2589 7 \\ 
Trappist-1 h & 2016 \citep{2017Natur.542..456G} & 2022 & NIRSpec & BOTS & CLEAR;PRISM & 1981 53 \\ 
\hline
GJ-9827 d & 2017 \citep{2017AJ....154..266N} & 2024 & NIRSpec & BOTS & F290LP;G395H & 4098 10 \\ 
\hline
TOI-776 b & 2019--2020 \citep{TOI776bDiscovery2021} & 2023 & NIRSpec & BOTS & F290LP;G395H & 2512 5 \\ 
TOI-776 c & 2019--2020 \citep{TOI776bDiscovery2021} & 2023 & NIRSpec & BOTS & F290LP;G395H & 2512 4 \\ 
\hline
GJ-357 b & 2019 \citep{GJ357bdiscovery2019} & 2023 & NIRSpec & BOTS & F290LP;G395H & 2512 7 \\ 
\hline
K2-18 b & 2014 \citep{2019ApJS..244...11K} & 2023 & NIRSpec & BOTS & F170LP;G235H & 2372 1 \\ 
\hline
L98-59 b & 2018--2019 \citep{2019AJ....158...32K} & 2024 & NIRSpec & BOTS & F290LP;G395H & 3942 4 \\ 
L98-59 c & 2018--2019 \citep{2019AJ....158...32K} & 2023 & NIRSpec & BOTS & F290LP;G395H & 2512 12 \\ 
L98-59 d & 2018--2019 \citep{2019AJ....158...32K} & 2024 & NIRSpec & BOTS & F290LP;G395H & 4098 2 \\ 
\hline
GJ-1132 b & 2014--2015 \citep{2015ESS.....310201B} & 2024 & NIRSpec & BOTS & F290LP;G395M & 1981 24 \\ 
\hline
LHS-1140 b & 2014--2015 \citep{2017Natur.544..333D} & 2023 & NIRSpec & BOTS & F290LP;G395H & 2334 1 \\ 
\hline
LP 791-18 c & 2015--2019 \citep{2019ApJ...883L..16C} & 2023 & NIRSpec & BOTS & CLEAR;PRISM & 1201 401 \\ 
\hline
TOI-270 b & 2015--2019 \citep{2019NatAs...3.1099G} & 2023 & NIRSpec & BOTS & F290LP;G395H & 4098 17 \\ 
TOI-270 d & 2015--2019 \citep{2019NatAs...3.1099G} & 2023 & NIRSpec & BOTS & F290LP;G395H & 4098 17 \\ 
\hline
TOI-178 b & 2018--2020 \citep{TOI178discovery2021} & 2023 & NIRSpec & BOTS & F290LP;G395M & 2319 3 \\ 
TOI-178 d & 2015--2019 \citep{TOI178discovery2021} & 2023 & NIRSpec & BOTS & F290LP;G395M & 2319 1 \\ 
TOI-178 g & 2015--2019 \citep{TOI178discovery2021} & 2023 & NIRSpec & BOTS & F290LP;G395M & 2319 102 \\ 
\hline
LTT-3780 c & 2019--2020 \citep{2020AJ....160....3C} & 2024 & NIRSpec & BOTS & F290LP;G395H & 3557 5 \\ 
\enddata
\end{deluxetable*}

\subsection{Data Reduction and white-light curve Production} \label{whitelight curveproduction}

Our light curves were produced using steps from Version 1.17.1 \citep{bushouse_jwst_2025} of the JWST Science Calibration Pipeline, which includes several packages that process the raw data into fully calibrated science data. We used the pipeline steps calwebb.detector 1, focusing on detector-level correction; calwebb.spec 2, which performs instrument-specific calibrations on the individual exposures; and calwebb.spec 3, which is responsible for the combination of the different exposures to obtain the final data product. For each target we retrieved the \texttt{rateints} products from MAST, which contain the uncalibrated non-destructive reads in the NRS1 and NRS2 detectors produced by the JWST Stage~1 pipeline. The reduction includes assignment of the World Coordinate System (WCS), correction for 1/\!$f$ noise, identification and interpolation of bad pixels, tracing of the spectral trace, and extraction of 1D spectra at each integration. The 1/\!$f$ noise and bad-pixel corrections were performed column-by-column by fitting a nominal spatial profile and replacing outliers beyond 3$\sigma$. Flat-fielding was omitted per \cite{espinoza_spectroscopic_2023}, who deduced based upon the instrumental stability that the effect of flat-fielding is negligible.  

Time-series observations gathered with NIRSpec are recorded simultaneously on two detectors (NRS1 and NRS2). Previous analyses that extracted white-light curves from both detectors have identified small but non-negligible detector-specific offsets. In particular, \citet{Sarkar24} report a $\sim$40–50 ppm difference in the recovered transit depth for NRS2, with the magnitude of the offset depending on whether the data were obtained in PRISM/CLEAR mode or with the F290LP/G395H grating for the same wavelength interval. Because our data set combines observations from both the PRISM and G395H configurations, we restrict our analysis to the NRS1 white-light curves to maintain a common transit-depth calibration across all epochs. Because NRS2, the redder detector, corresponds to wavelengths $>3\mu$m (longer than the peak wavelength for even the coolest star in our sample, TRAPPIST-1), the majority of the stellar flux falls onto NRS1 in any case. The photometric error for NRS2 is corresponding higher, reflecting an underlying increase in photometric uncertainty moving from shorter to longer NIRSpec wavelengths \citep{Sarkar24, Gordon25}. We performed one case study for GJ~9827 d, separately extracting and fitting the light curves for NRS1 and NRS2; we found that the duration uncertainty is 50\% higher for the NRS2 detector in this case.

It is worth a brief note to consider the validity of calculating \textit{only} the white-light transit duration. If the transit depth is changing due to atmospheric absorption (the ideal situation for atmospheric study), then the larger/smaller planet silhouette will slightly alter the timing of first and fourth contact, the points that bracket the interval we are measuring. As a test, we consider the case study of a $\Delta\delta=40$ ppm difference in transit depth between two wavelengths. For a 0.2$M_{\odot}$ M dwarf with a transiting $1.5R_{\oplus}$ planet (corresponding to a $\sim$2500 ppm transit), a 40 ppm depth difference corresponds to the planet's apparent radius changing by 0.8\%. To first order, the duration of ingress/egress is $\sim R_{p}/v$, where $v$ is the orbital velocity: 2-3 minutes for a typical planet in our sample. An apparent change in planet radius of 0.8\% corresponds to a difference in ingress/egress duration of 1-2 seconds. As we show below, our typical $\Delta\tau$ for the JWST transits is 5-10 seconds. For this reason, we have ignored this effect for the purposes of this study; a more sophisticated analysis will be possible once all transmission spectra for these data are published. 

After producing the white-light curves, we removed low-order systematics by fitting and dividing out a polynomial baseline to the out-of-transit flux. JWST NIRSpec time-series data often exhibit a slight linear trend across each integration, so baseline correction is a standard step before transit fitting. Because we model the transit light curves using the \texttt{batman} package \citep{kreidberg_batman_2015}, which assumes multiplicative systematics have already been removed, we perform this de-trending prior to the MCMC analysis. For each planet’s white-light curve, we fit a second-degree polynomial to the off-transit flux using \texttt{numpy.polyfit}, and normalize the light curve by dividing through by this best-fit baseline. The uncertainties on the normalized flux values were estimated from the scatter of the off-transit data after the baseline correction. This procedure yields a more stable estimate of the mean out-of-transit flux and results in more reliable uncertainty estimates for the subsequent transit fitting.

\subsection{Duration Measurements} \label{subsec:whitelight curvefitting}

For light curve fitting, we employ \texttt{batman} \citep{kreidberg_batman_2015} within a Monte Carlo Markov Chain (MCMC) algorithm included in the emcee \citep{emcee} package. In addition to the planet-star radius ratio (${R_p}/{R_\star}$), we fit for mid-transit time ($T_0$), the orbital period $P$, the ratio of semi-major axis to star radius $a/R_{\star}$, the impact parameter $b$, and two quadratic limb darkening coefficients (LDCs). We employed a Gaussian likelihood function, with per-measurement uncertainty set by the out-of-transit standard deviation. We experimented with allowing the period to float, versus fixing its value; in practice, fixing the orbital period to within the measurement error did not substantially reduce the duration uncertainty. With respect to limb darkening, we allowed both parameters to float, rather than employing the model LDCs that exist specific to NIRSpec from the Exoplanet Characterization ToolKit\footnote{https://exoctk.stsci.edu/limb\_darkening}. \cite{Sarkar24} found that photometric precision was improved for empirical quadratic LDC fits, as opposed to model LDCs. We employ the model fit posterior chains to compute the duration per Equation \ref{eq:duration}. We fix eccentricity to be zero, reasonable for M dwarf multis per \cite{sagear_orbital_2023}. 

\begin{equation}
T_{tot} = t_{IV} - t_{I} = \frac{P}{\pi} \cdot \arcsin \left( 
\frac{R_*}{a} \cdot 
\frac{\sqrt{\left( 1 + \frac{R_p}{R_*} \right)^2 - b^2}}{\sin i} 
\right)
\label{eq:duration}
\end{equation}

\begin{figure*}
\gridline{
  \fig{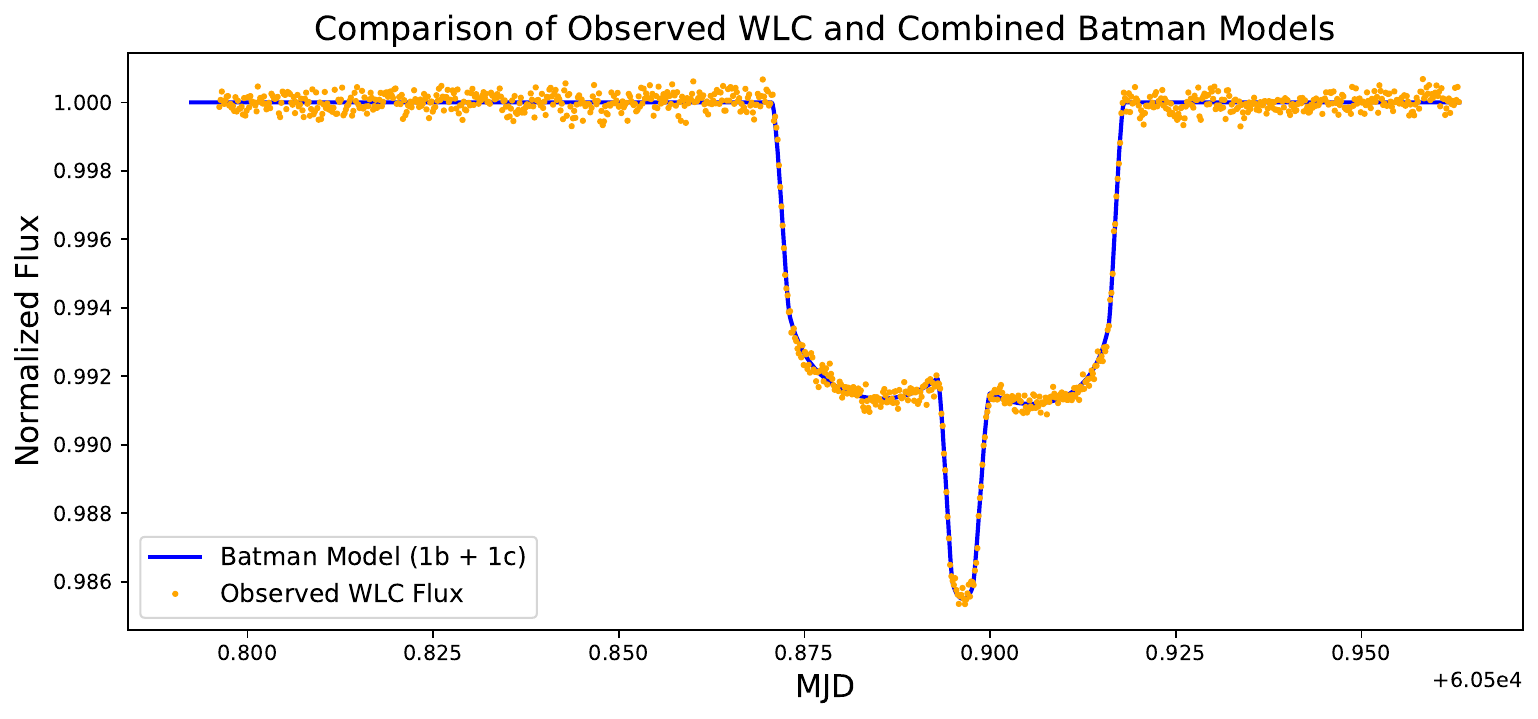}{0.5\textwidth}{}
  \fig{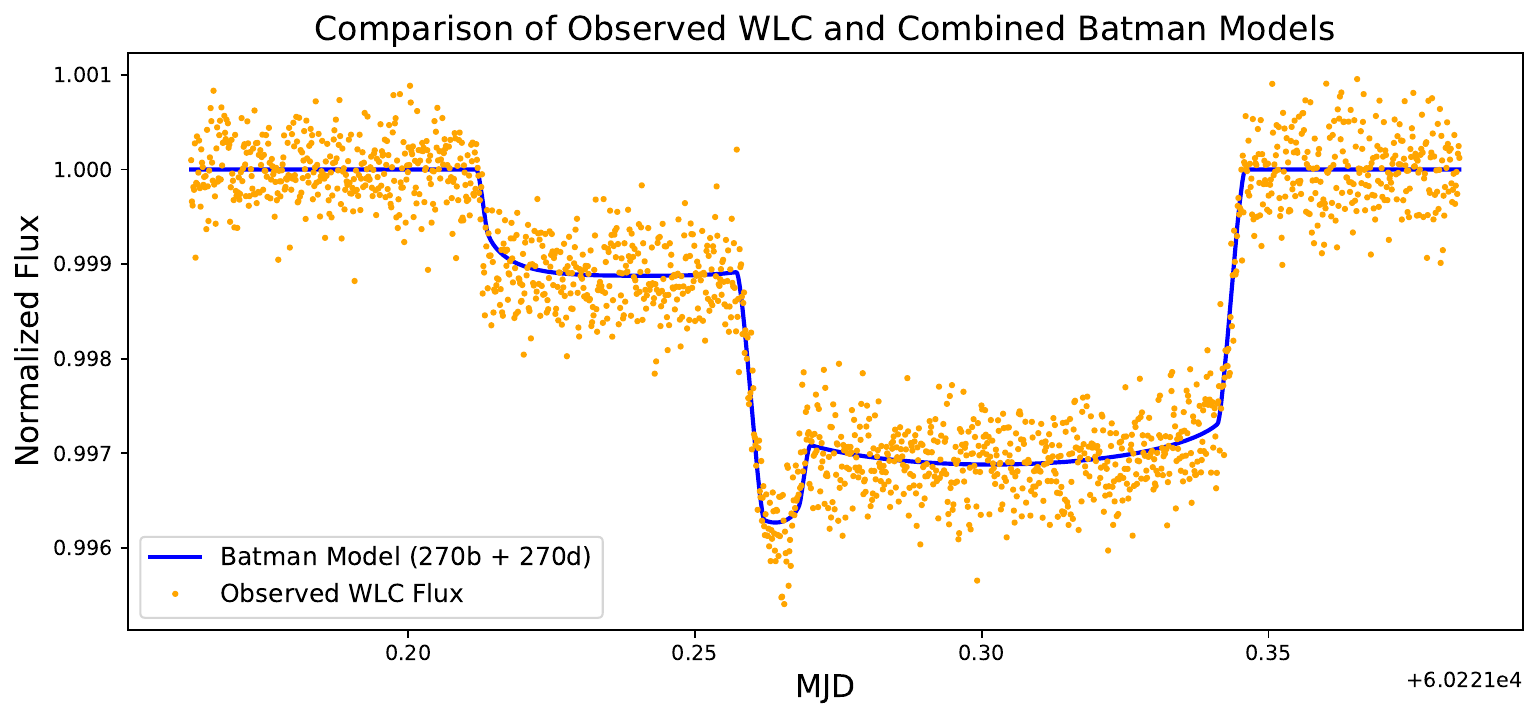}{0.5\textwidth}{}
}
\caption{\textit{Left panel:} white-light curve of TRAPPIST-1b and TRAPPIST-1c during a simultaneous transit event, with best-fit model over-plotted. The transit of TRAPPIST-1c occurred first, followed by TRAPPIST-1b. \textit{Right:} white-light curve of TOI-270b and TOI-270d during a simultaneous transit event, with best-fit model over-plotted. The transit of TOI-270b occurred first, followed by TOI-270d.}
\label{fig:wlc}
\end{figure*}

We note that our sample contains two JWST observations of simultaneous transits: TRAPPIST-1b and TRAPPIST-1c, and TOI-270b and TOI-270d. 
We illustrate our white-light–curve fitting result in Figure \ref{fig:wlc} for these two observations. In these cases, we constructed a light curve model that included the two transits together (varying both sets of transit parameters simultaneously), evaluating their joint likelihood in the MCMC fit. Table \ref{table:parametersfortransitdurationchange} contains the measured transit durations for the JWST observations in our sample. 

\section{Analysis} \label{sec:analysis}

In this Section, we consider the JWST transit durations against literature values. First, we fit a linear relationship between the JWST duration and published durations from the literature for each exoplanet. This procedure is described in Section \ref{subsec:timeevolutionoftransitduration}. In Section \ref{subsec:comparisonbetweentheoreticalprediction}, we model to first order the prediction for TDVs in each system, assuming nominal values for $g$ and marginalizing over uncertainties in the transit configuration. We then compare the observed transit duration variation for each system against our resulting predictions. 

\subsection{Evaluation of Transit Duration Variation} \label{subsec:timeevolutionoftransitduration}

To identify other published transit duration measurements for each plane, we employ the python package \texttt{astroquery}. The relevant manuscripts for each planet are summarized in Table \ref{table:planetsampletimeevolution}. We have prioritized datasets gathered from space, in  consideration of the generally higher precision. By searching each manuscript, we identified the epoch of the transit corresponding to the duration measurement. 

Because the precession timescale is assumed to be much longer than the observation baseline \citep{Millholland21}, we employ a linear fit as a reasonable approximation to the transit duration change over time. In this case, $\tau=m\cdot{\tau_0}+b$, where $\tau_0$ is the discovery epoch and the slope parameter $m$ is in minutes yr$^{-1}$. We performed an MCMC fit to the set of duration measurements for each planet, used their reported error bars and varying slope and intercept. We assumed that the reported duration errors were normally distributed, and so employed a Gaussian likelihood function. Where the published duration measurement corresponded to the average over many transits, we employed the mean of the baseline as the transit epoch. We identified the best-fit slope by the mode of the distribution and the 1$\sigma$ interval from the closest 68\% of samples to this mean value. For each slope and intercept, we evaluate the model line at the epochs of the first and last transit measurements for each planet, recording this $\Delta\tau$ value.  In Table \ref{table:parametersfortransitdurationchange}, we show the mean and $1\sigma$ confidence interval on $\Delta \tau$. We also report $\Delta\tau/\sigma_{\Delta\tau}$, the best-fit slope normalized by our uncertainty in that slope. 

\begin{figure*}
\centering
\gridline{
  \fig{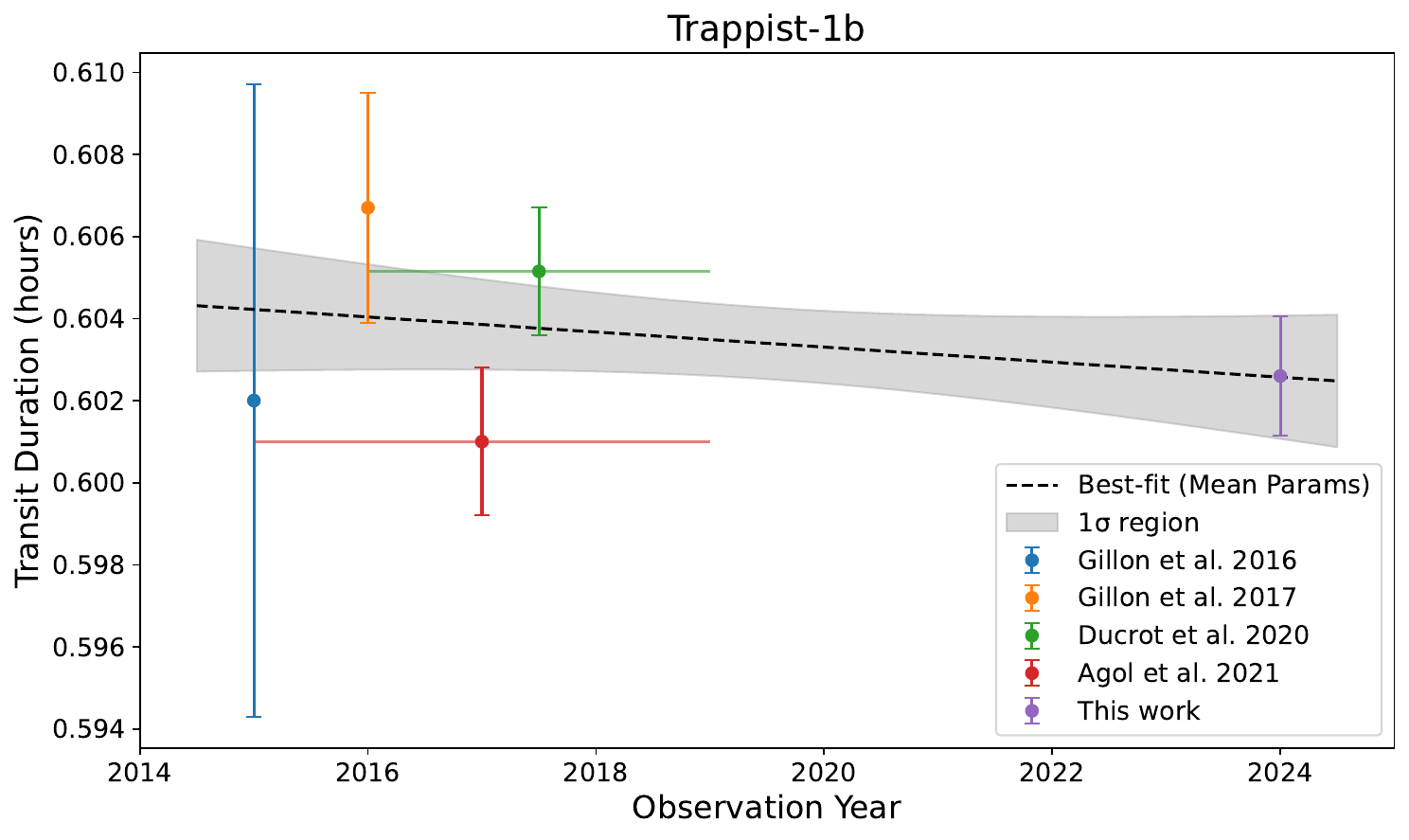}{0.33\textwidth}{}
  \fig{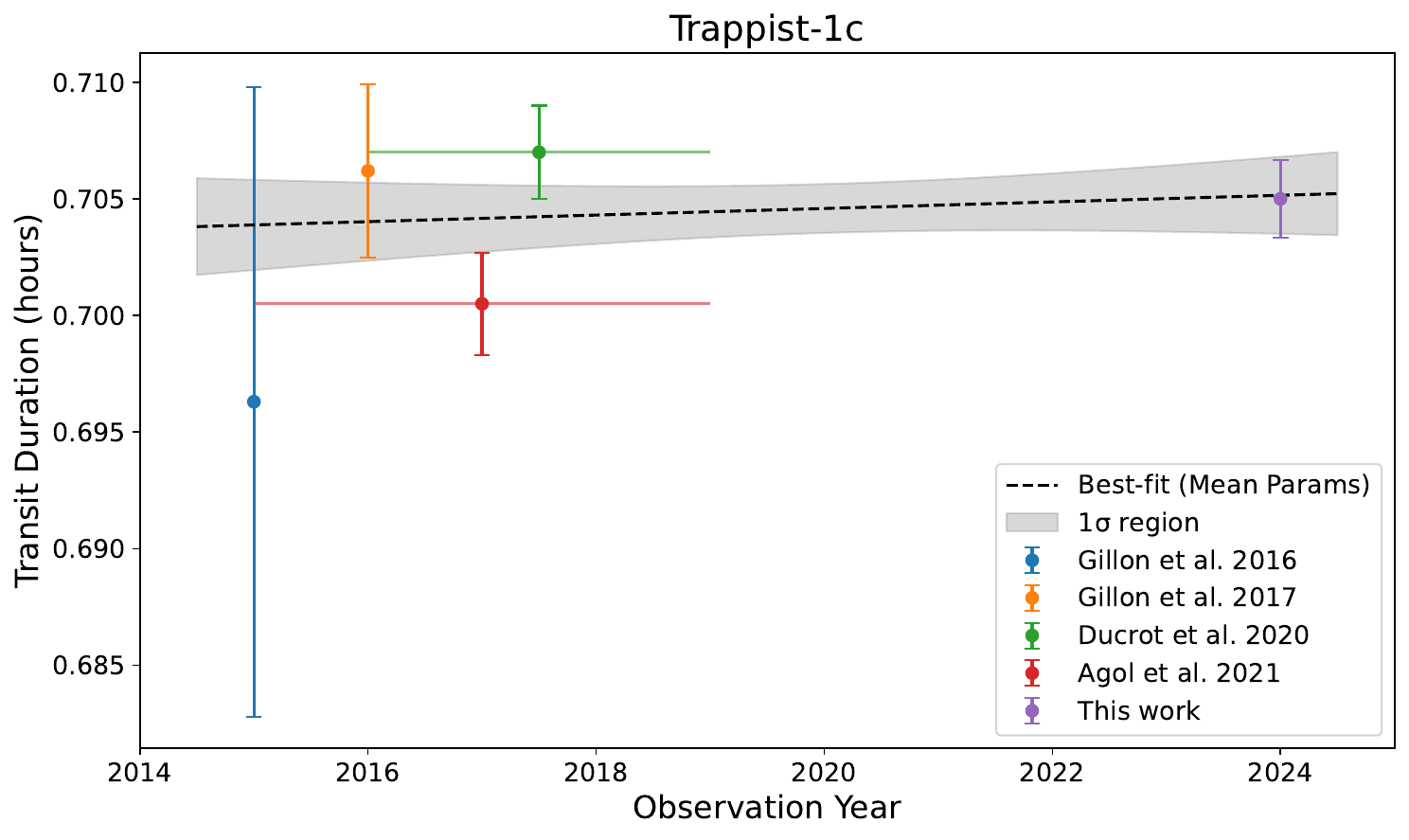}{0.33\textwidth}{}
  \fig{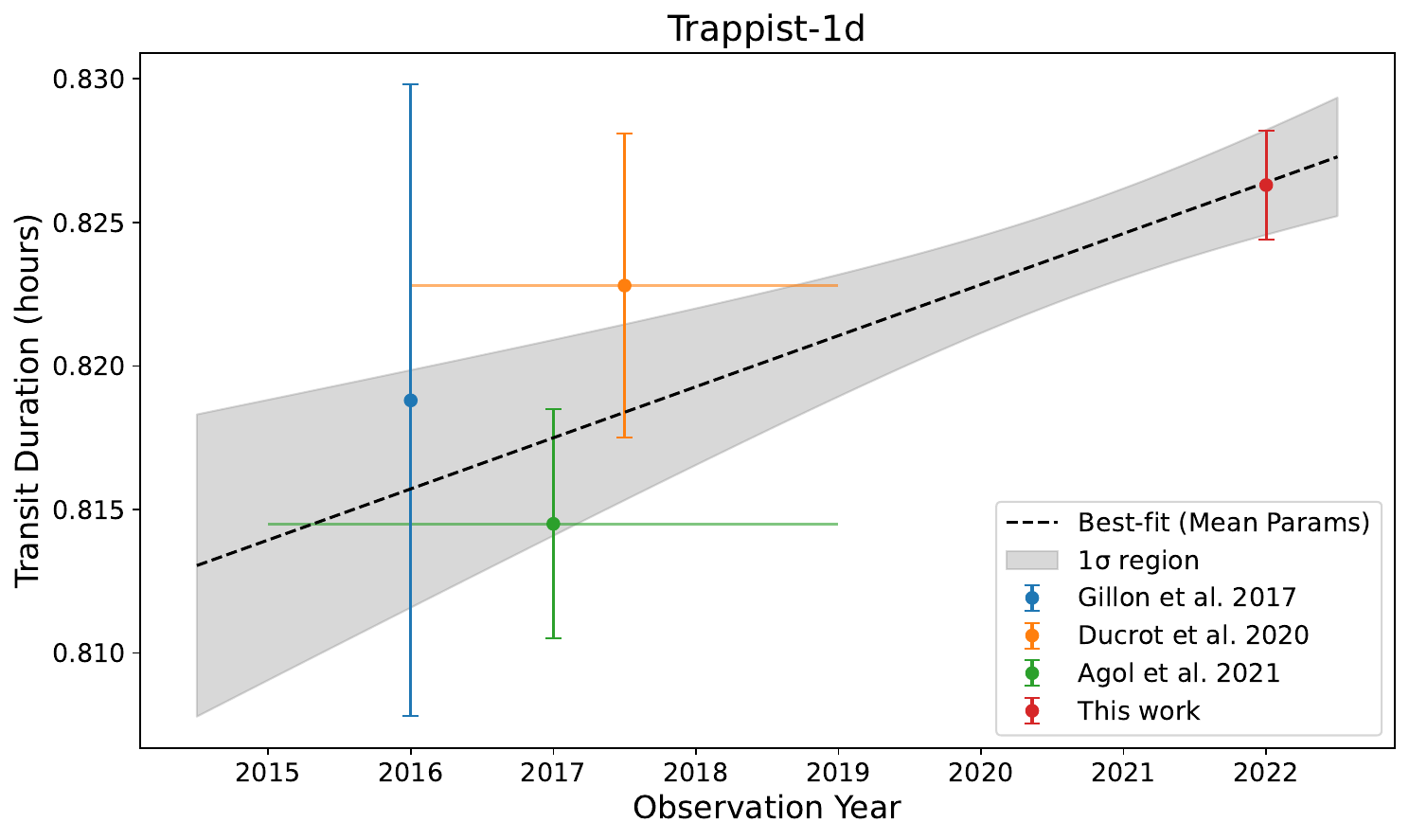}{0.33\textwidth}{}
}
\vspace{-20pt}
\gridline{
  \fig{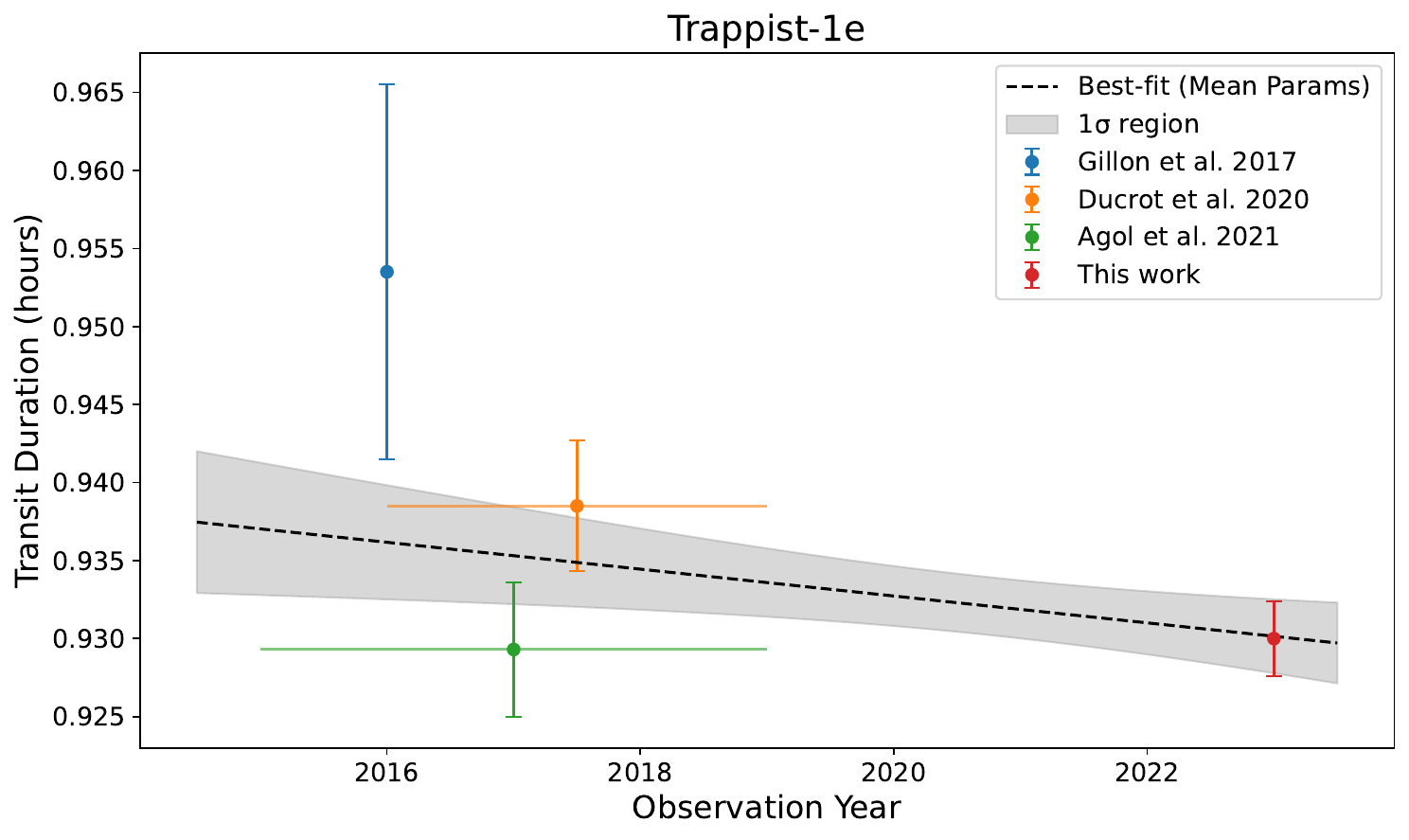}{0.33\textwidth}{}
  \fig{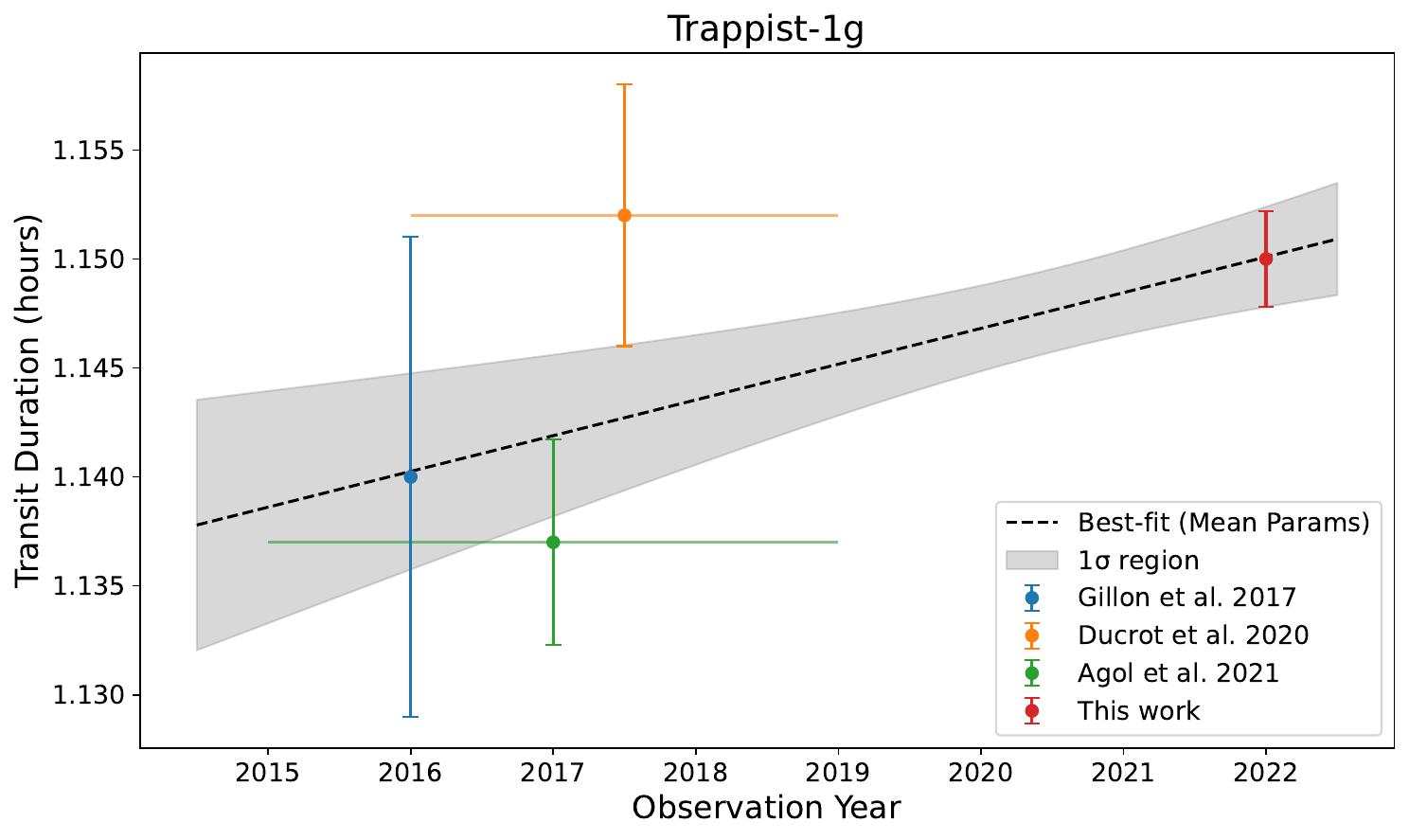}{0.33\textwidth}{}
  \fig{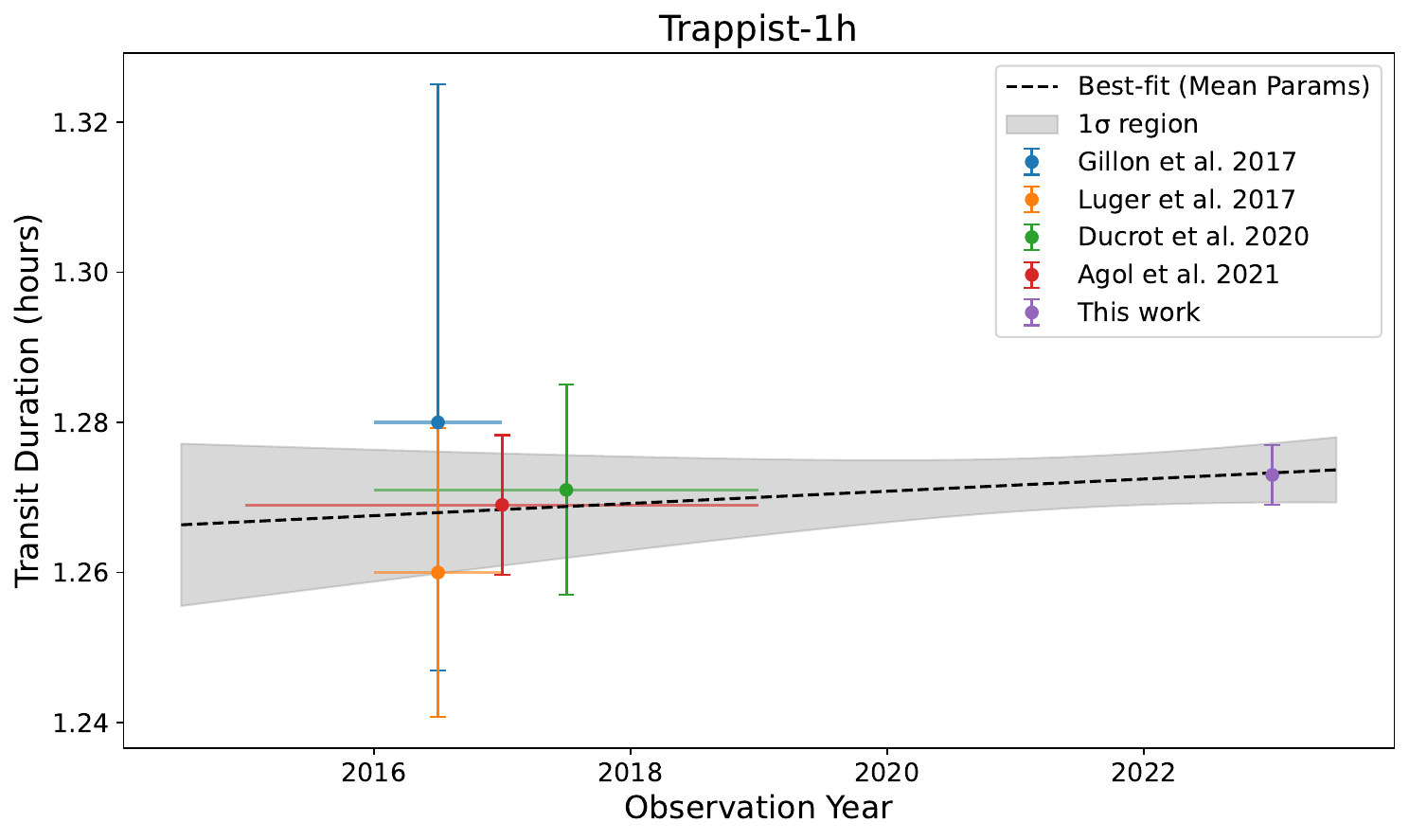}{0.33\textwidth}{}
}

\caption{Time evolution of transit duration for the TRAPPIST-1 system.}
\label{fig:TDV_TRAPPIST1}
\end{figure*}

\begin{figure*}
\centering
\gridline{
  \fig{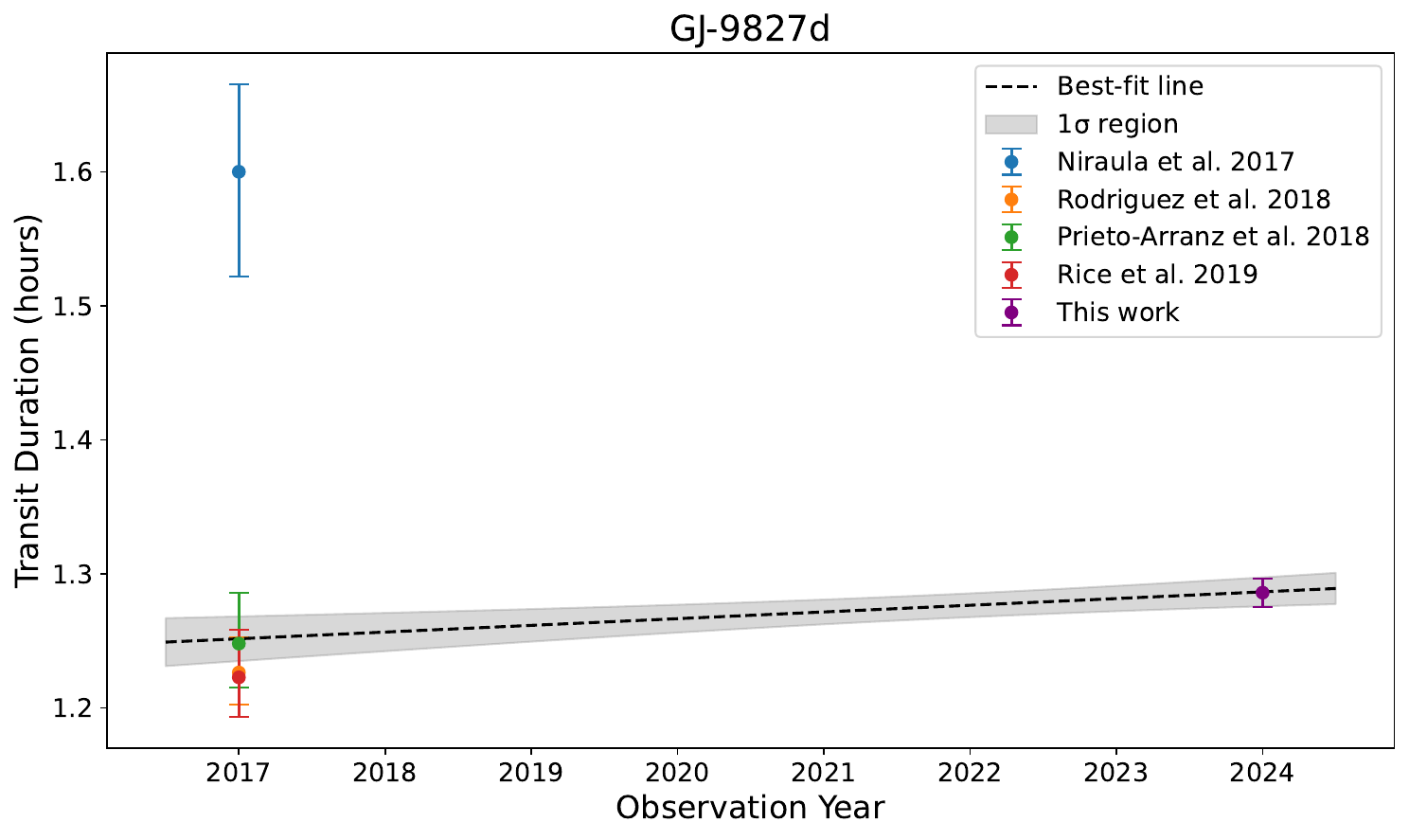}{0.33\textwidth}{}
  \fig{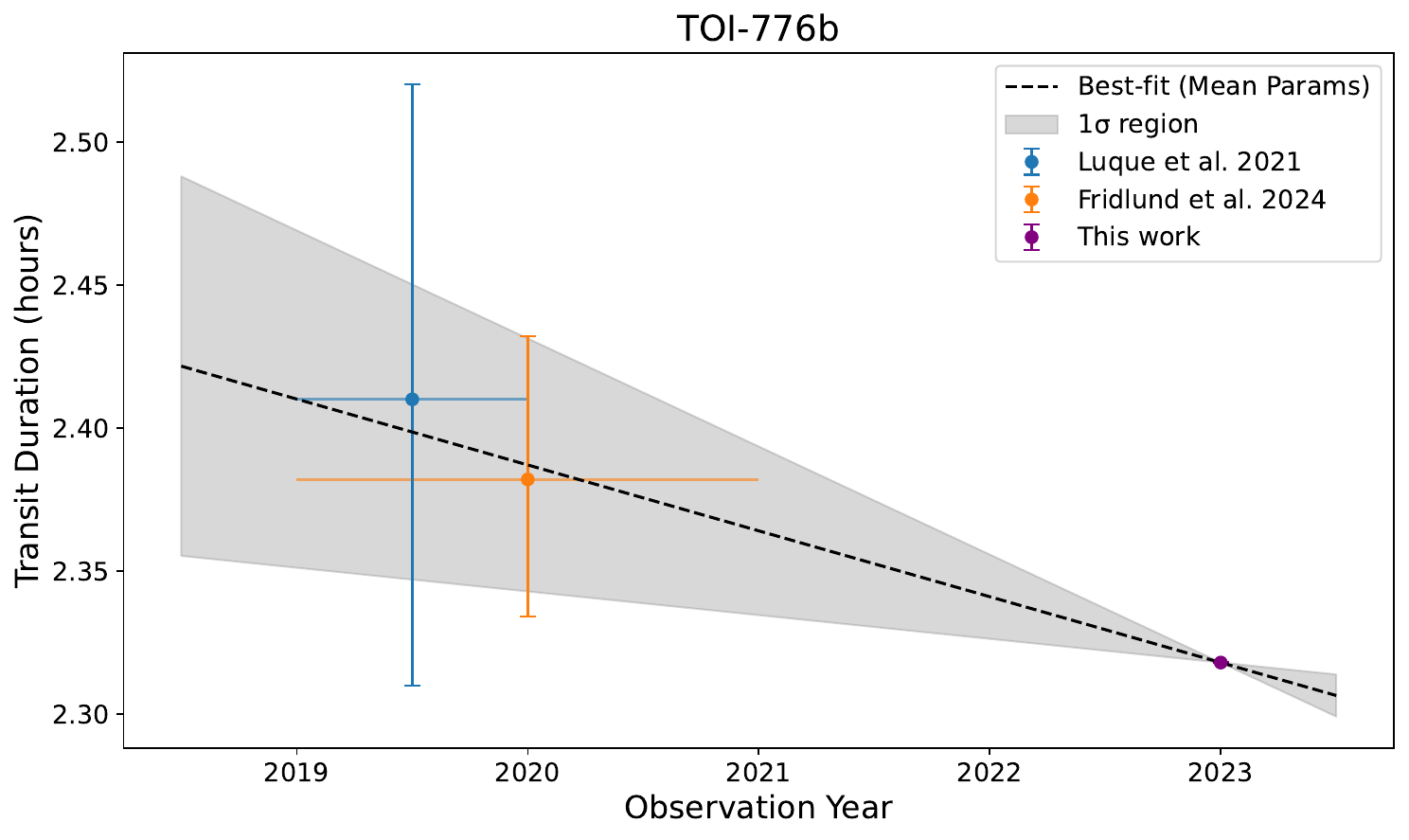}{0.33\textwidth}{}
  \fig{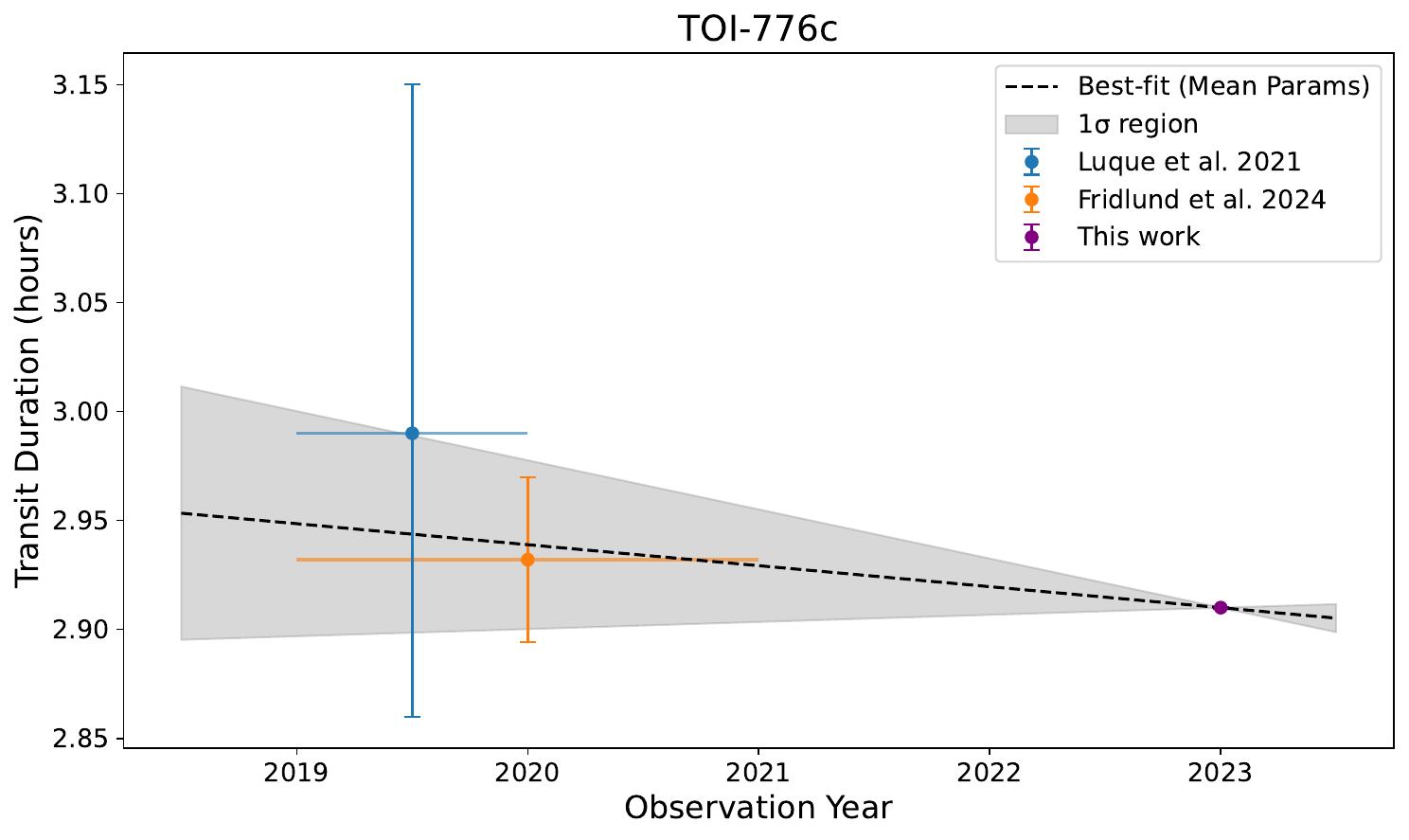}{0.33\textwidth}{}
}
\vspace{-20pt}
\gridline{
  \fig{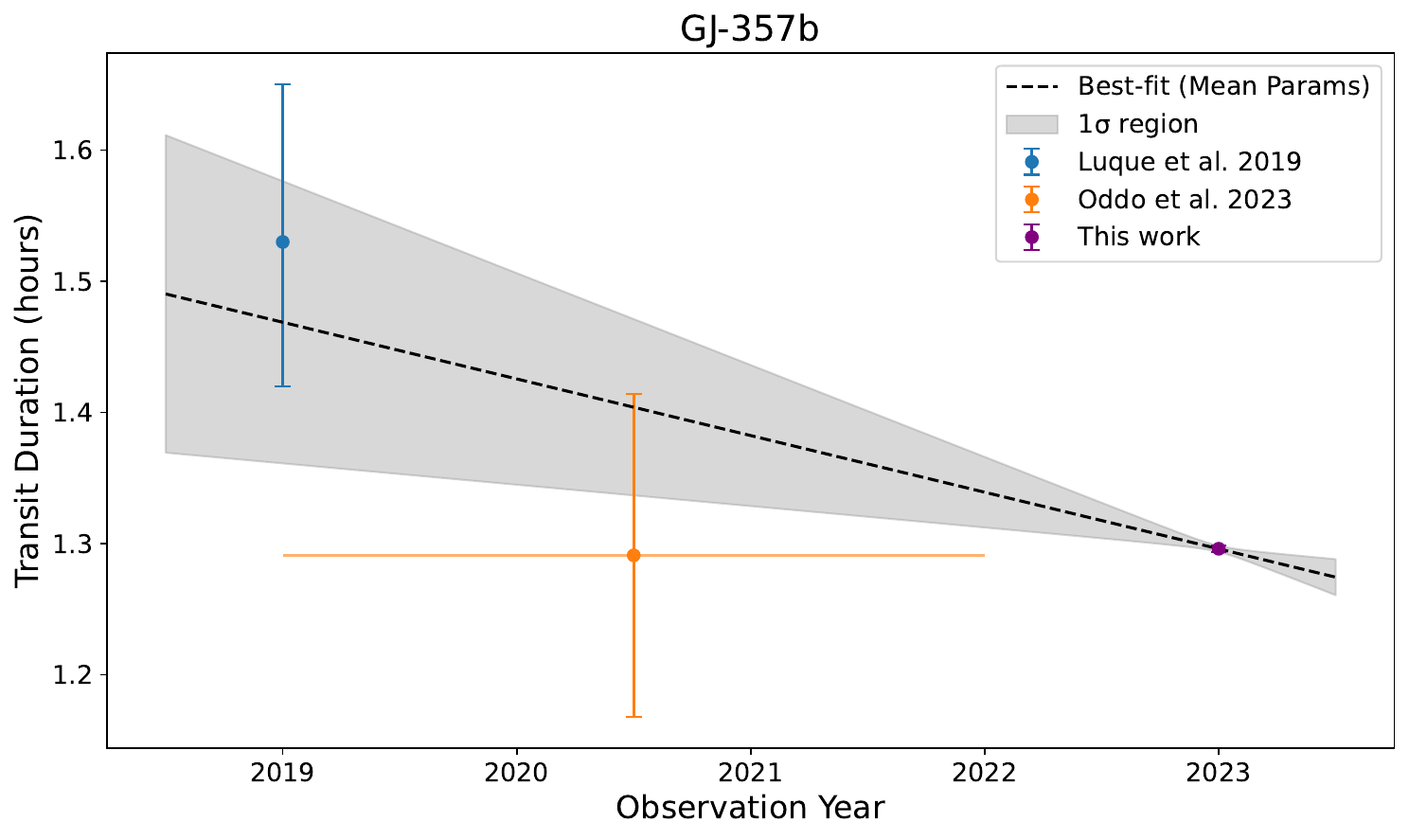}{0.33\textwidth}{}
  \fig{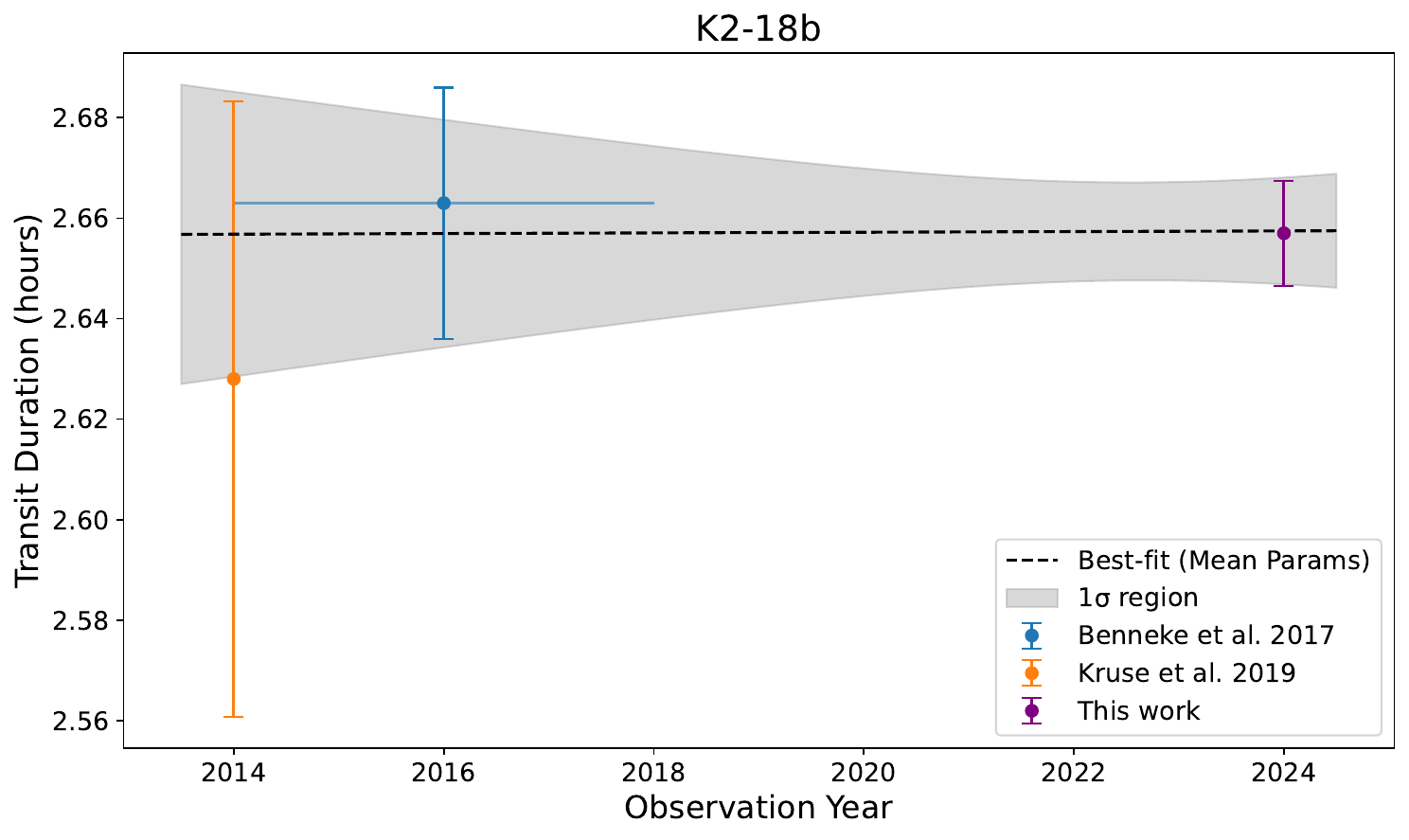}{0.33\textwidth}{}
  \fig{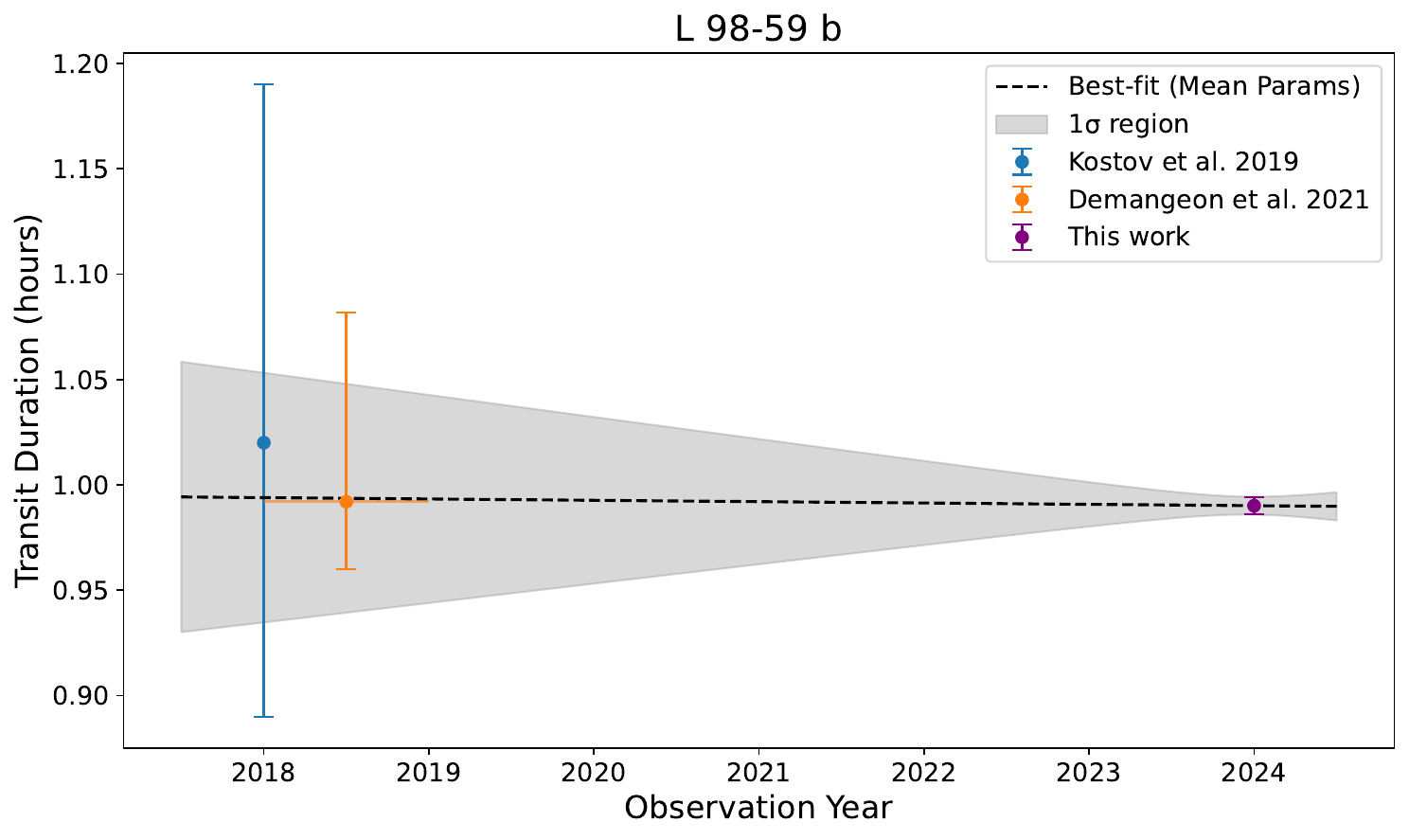}{0.33\textwidth}{}
}

\caption{Time evolution of transit duration for GJ-9827, TOI-776, GJ-357, K2-18, and L98-59.}
\label{fig:TDV_GJ_TOI_K2_L9859b}
\end{figure*}

\begin{figure*}
\centering
\gridline{
  \fig{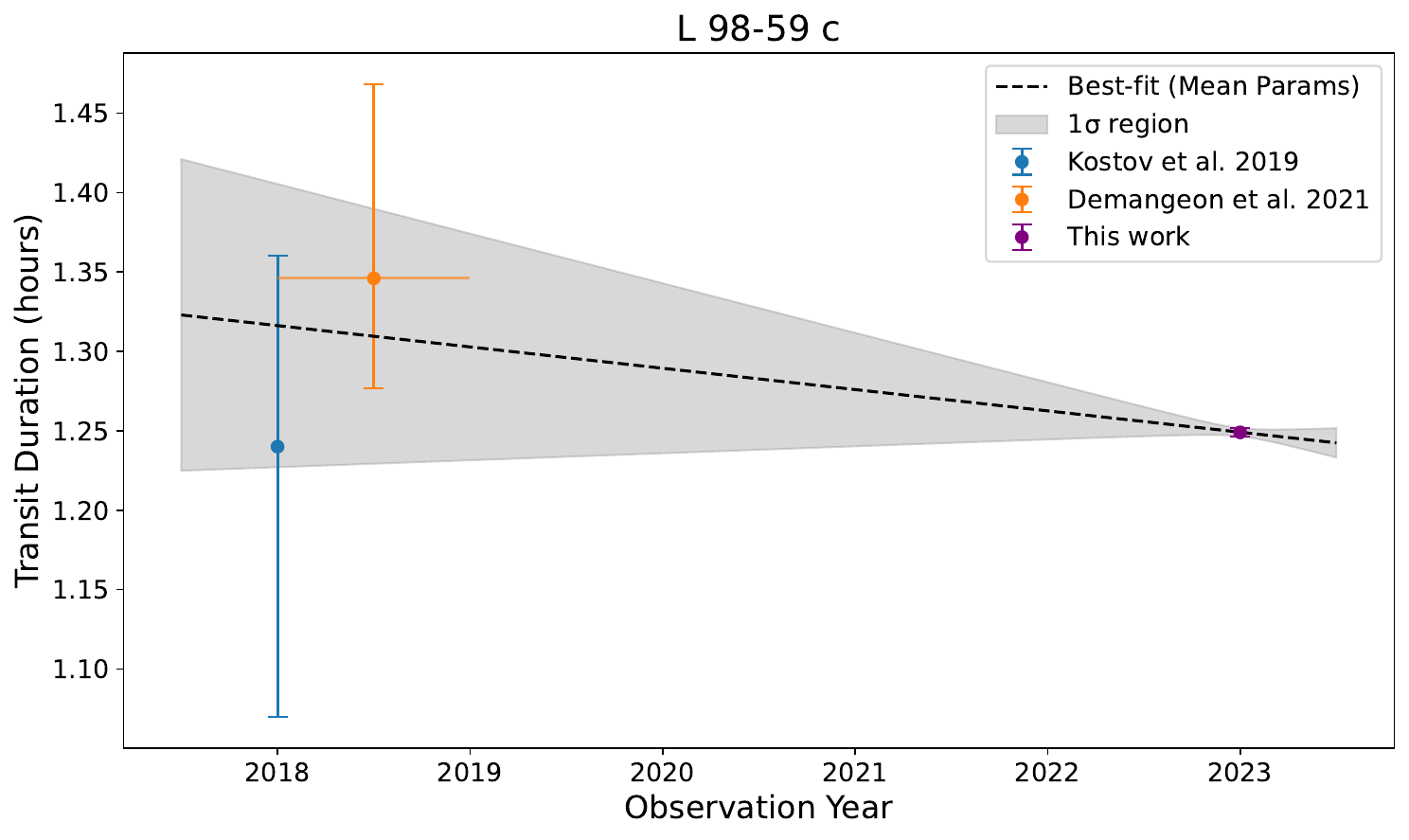}{0.33\textwidth}{}
  \fig{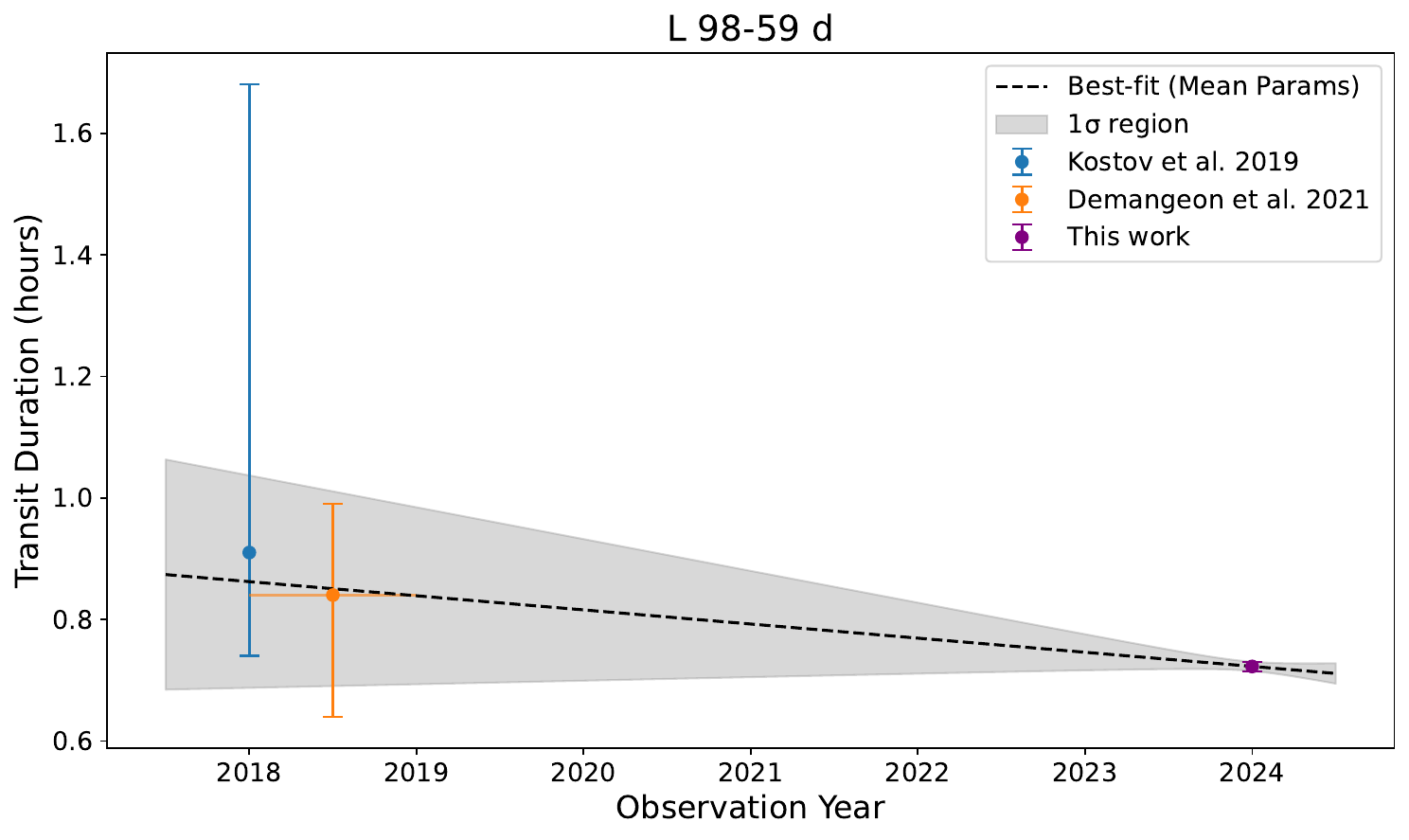}{0.33\textwidth}{}
  \fig{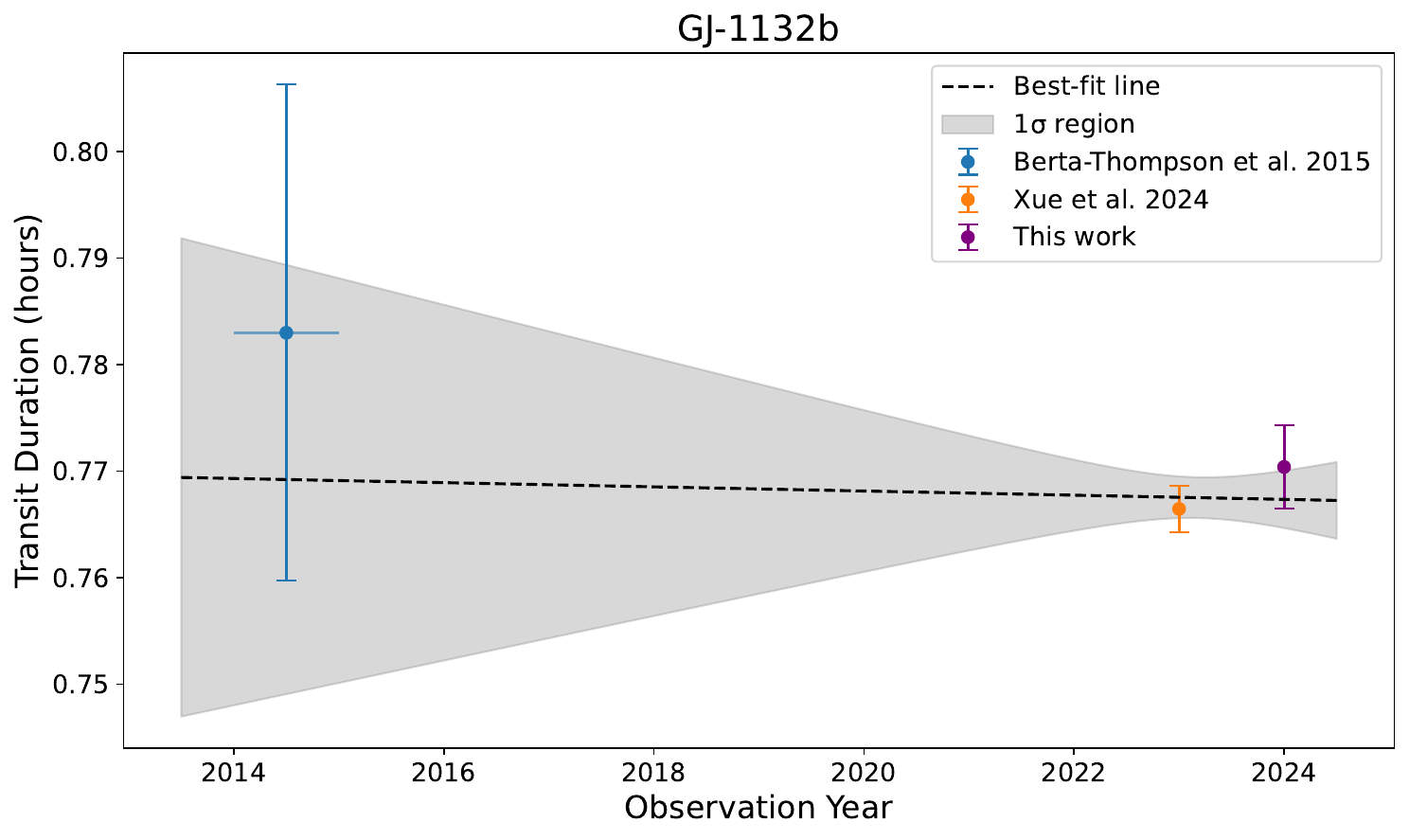}{0.33\textwidth}{}
}
\vspace{-20pt}
\gridline{
  \fig{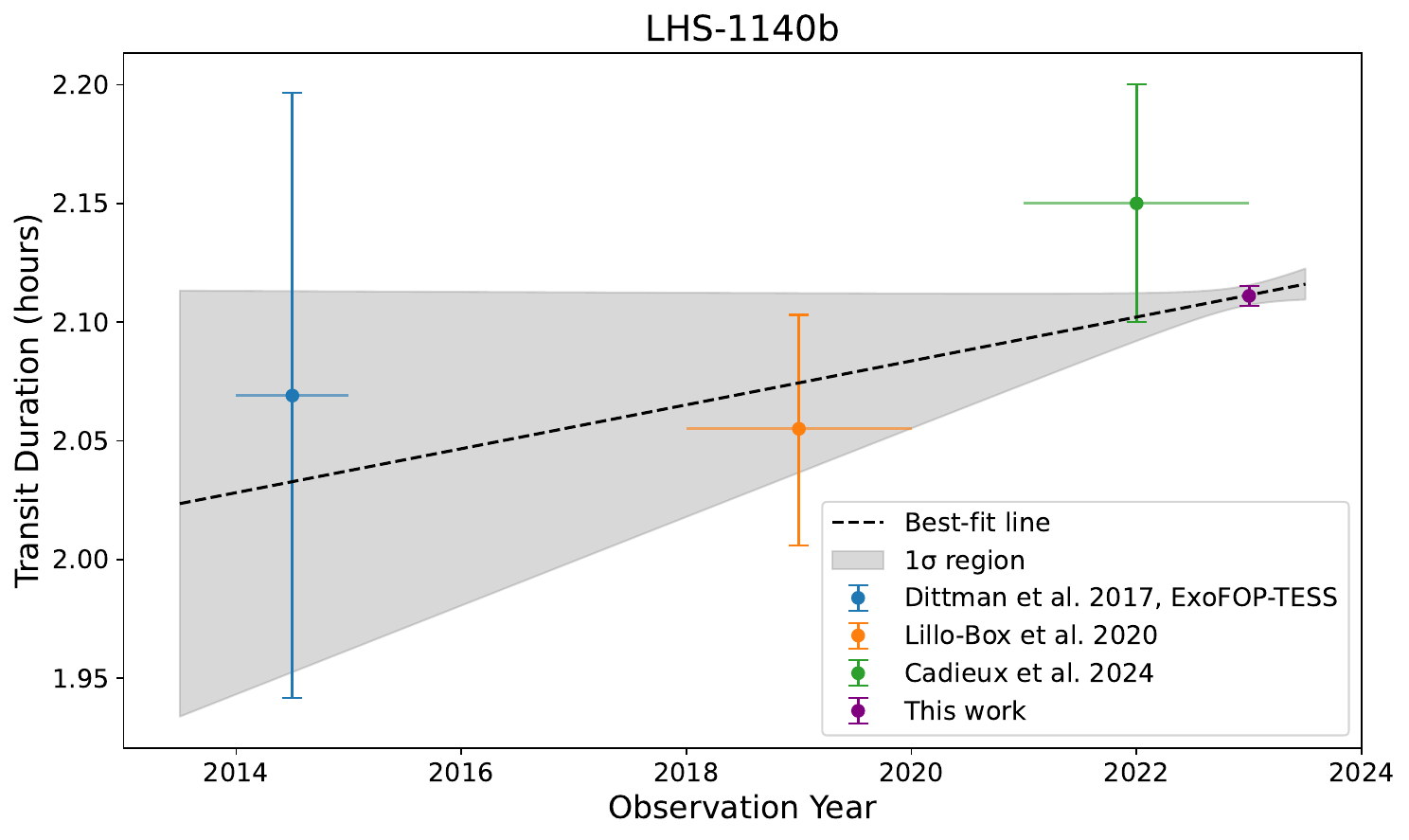}{0.33\textwidth}{}
  \fig{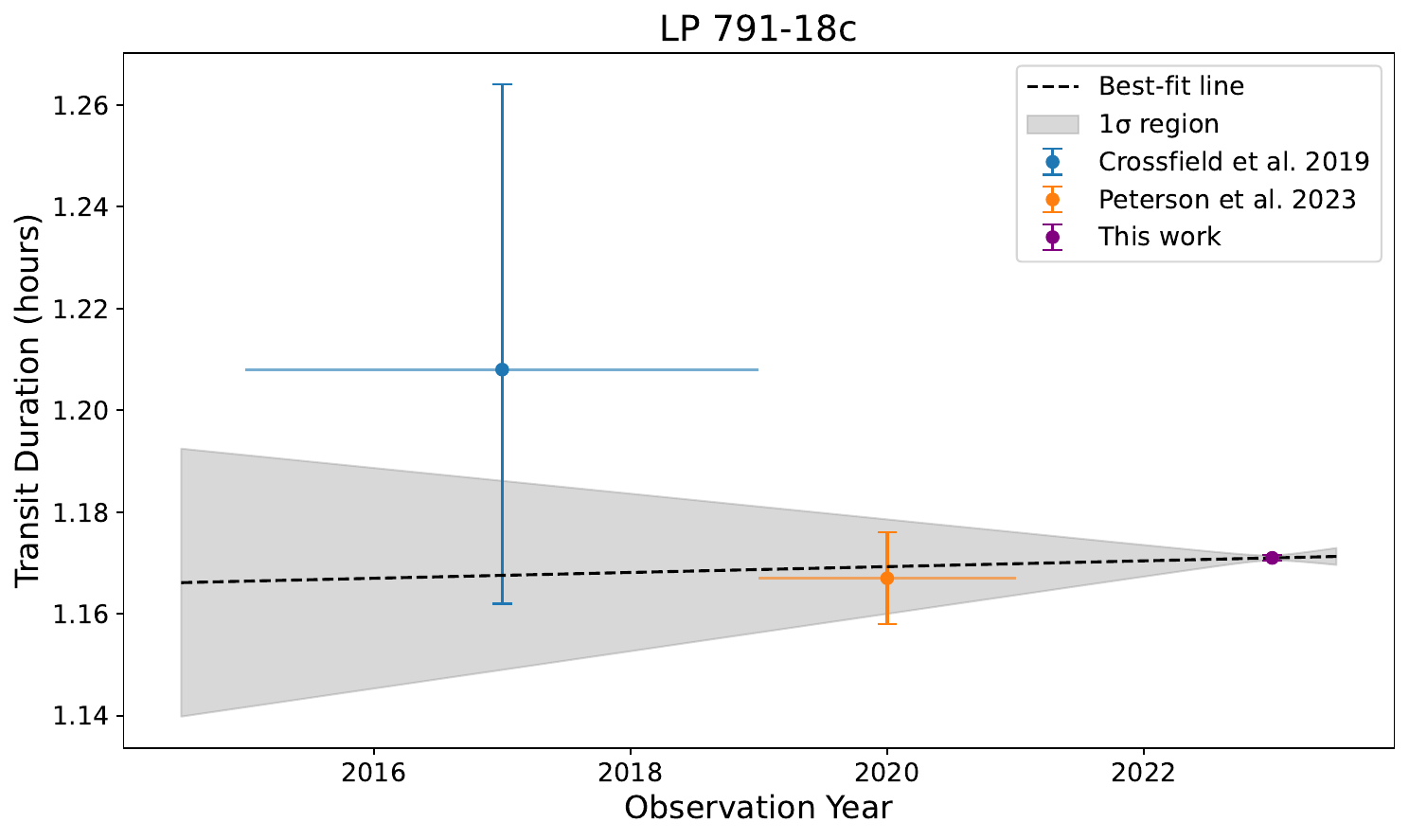}{0.33\textwidth}{}
  \fig{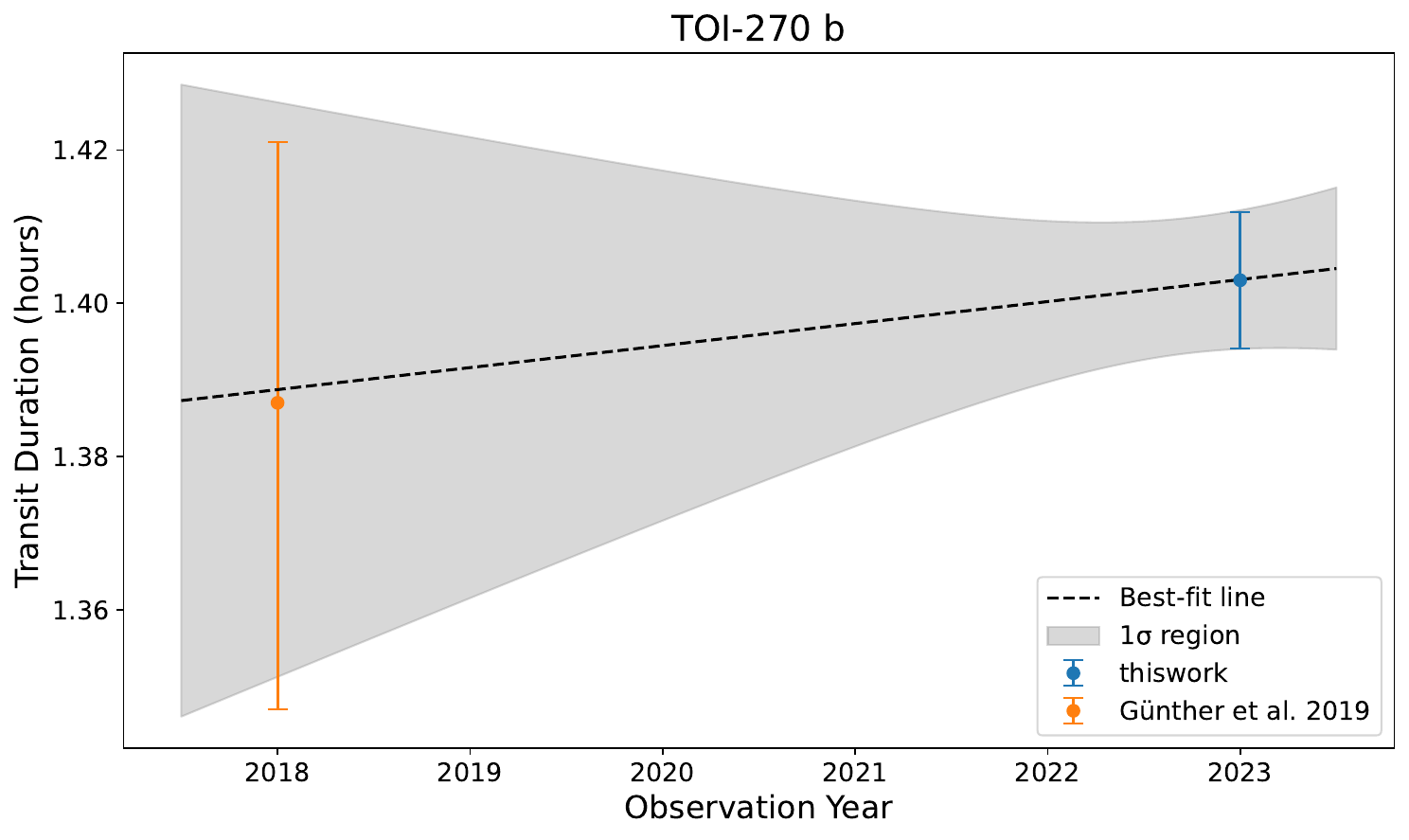}{0.33\textwidth}{}
}
\vspace{-20pt}
\gridline{
  \fig{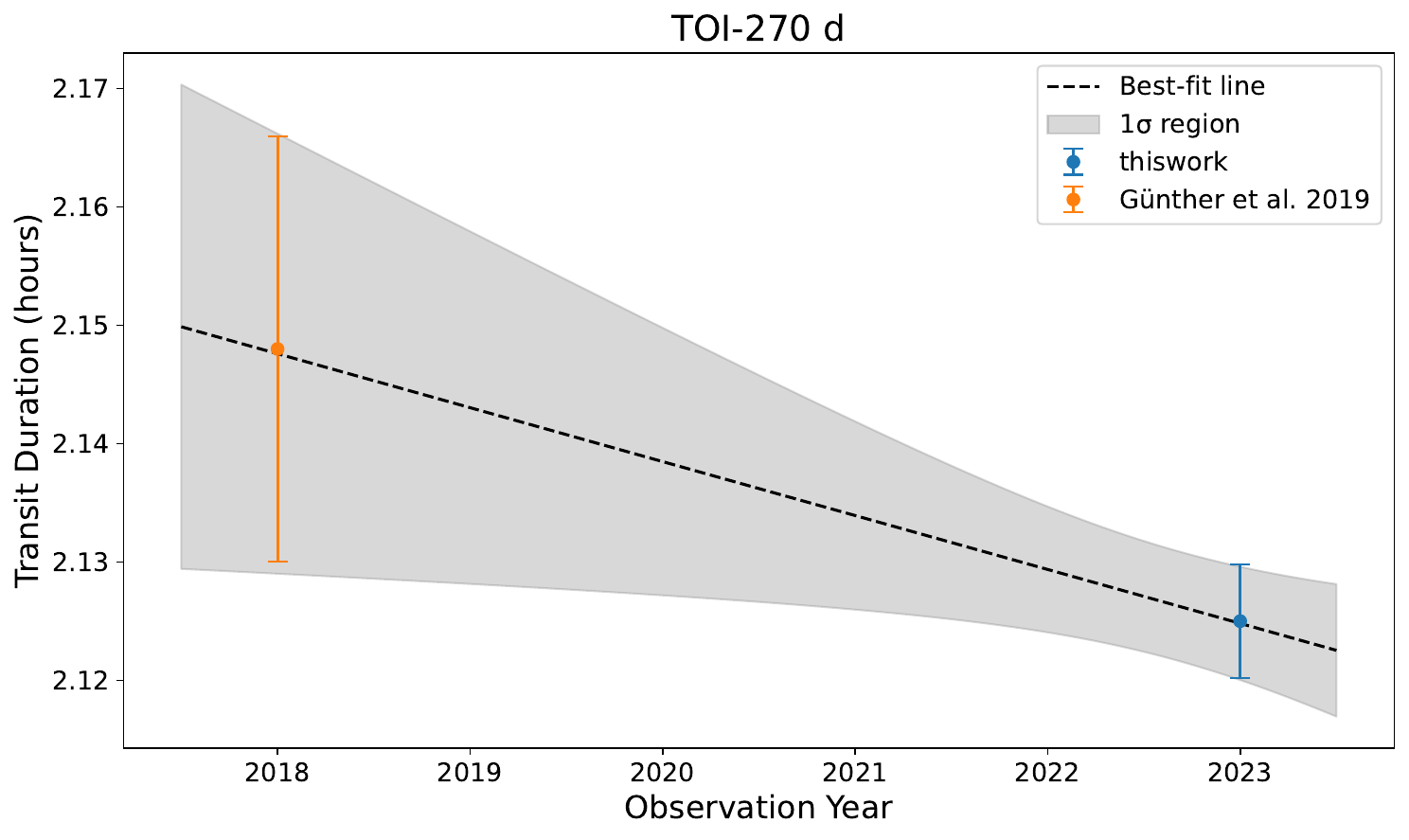}{0.33\textwidth}{}
  \fig{paperplots/Transit_Duration_Time_evolution_TOI-178b.pdf}{0.33\textwidth}{}
  \fig{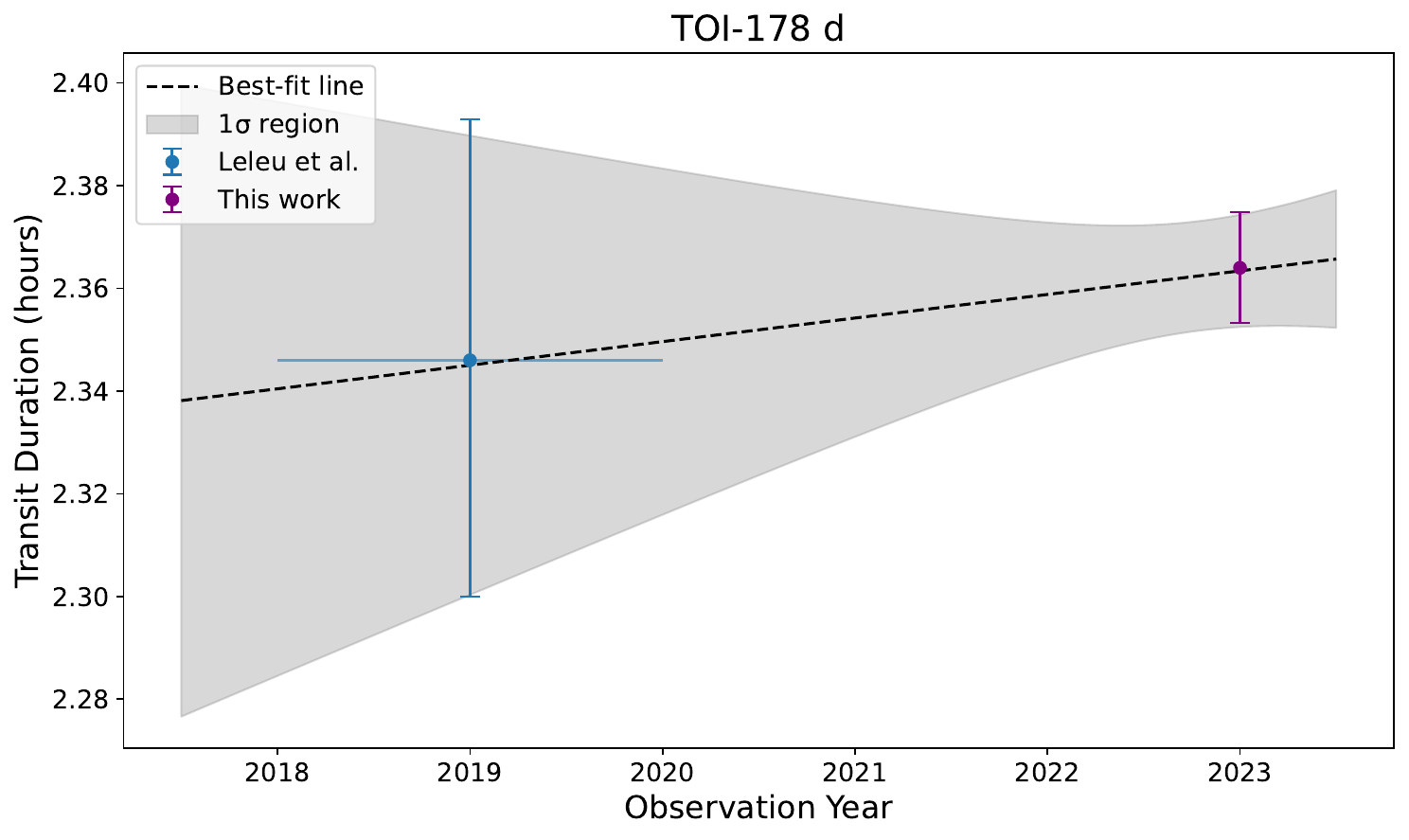}{0.33\textwidth}{}
}
\vspace{-20pt}
\gridline{
  \hfill
  \fig{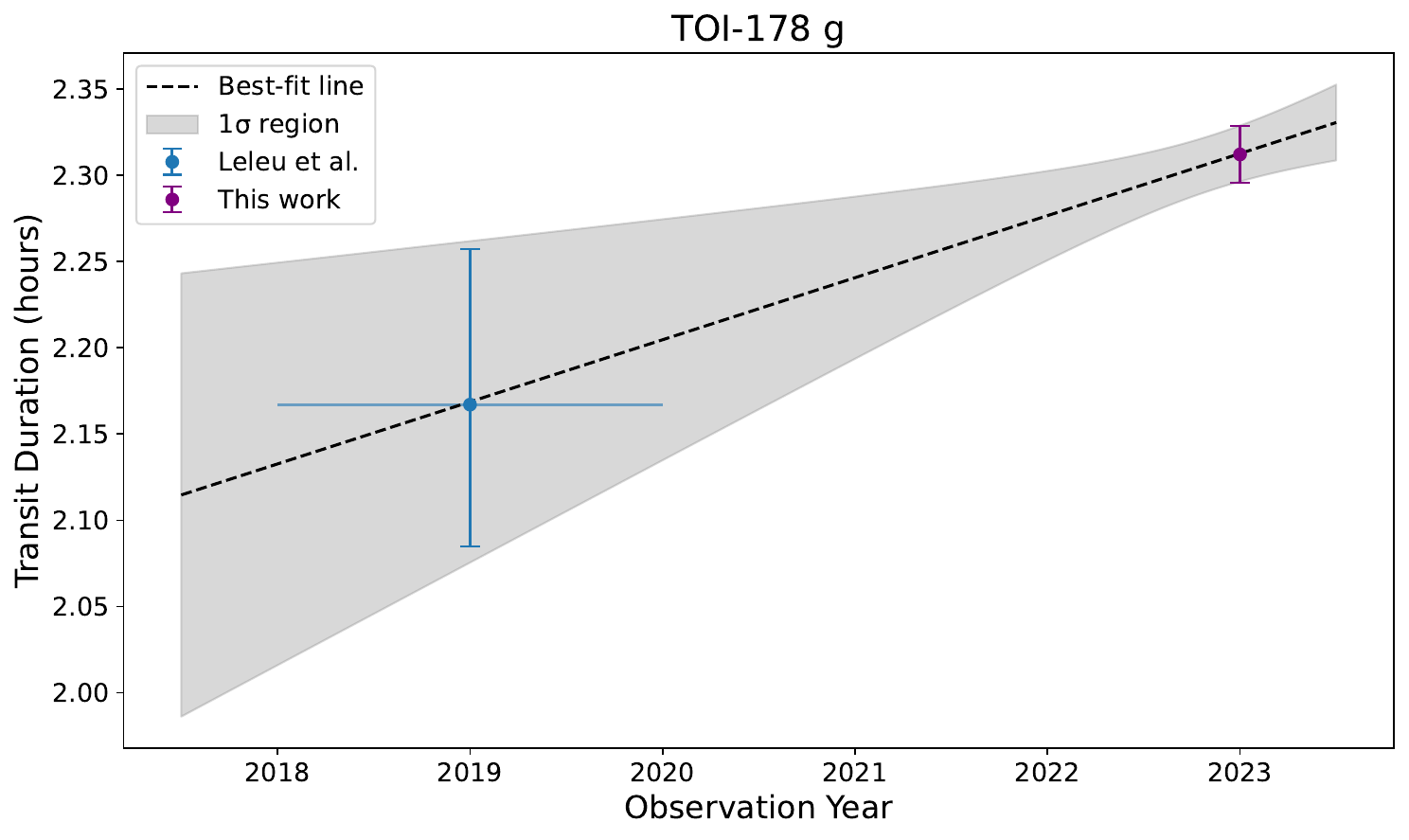}{0.33\textwidth}{}
  \fig{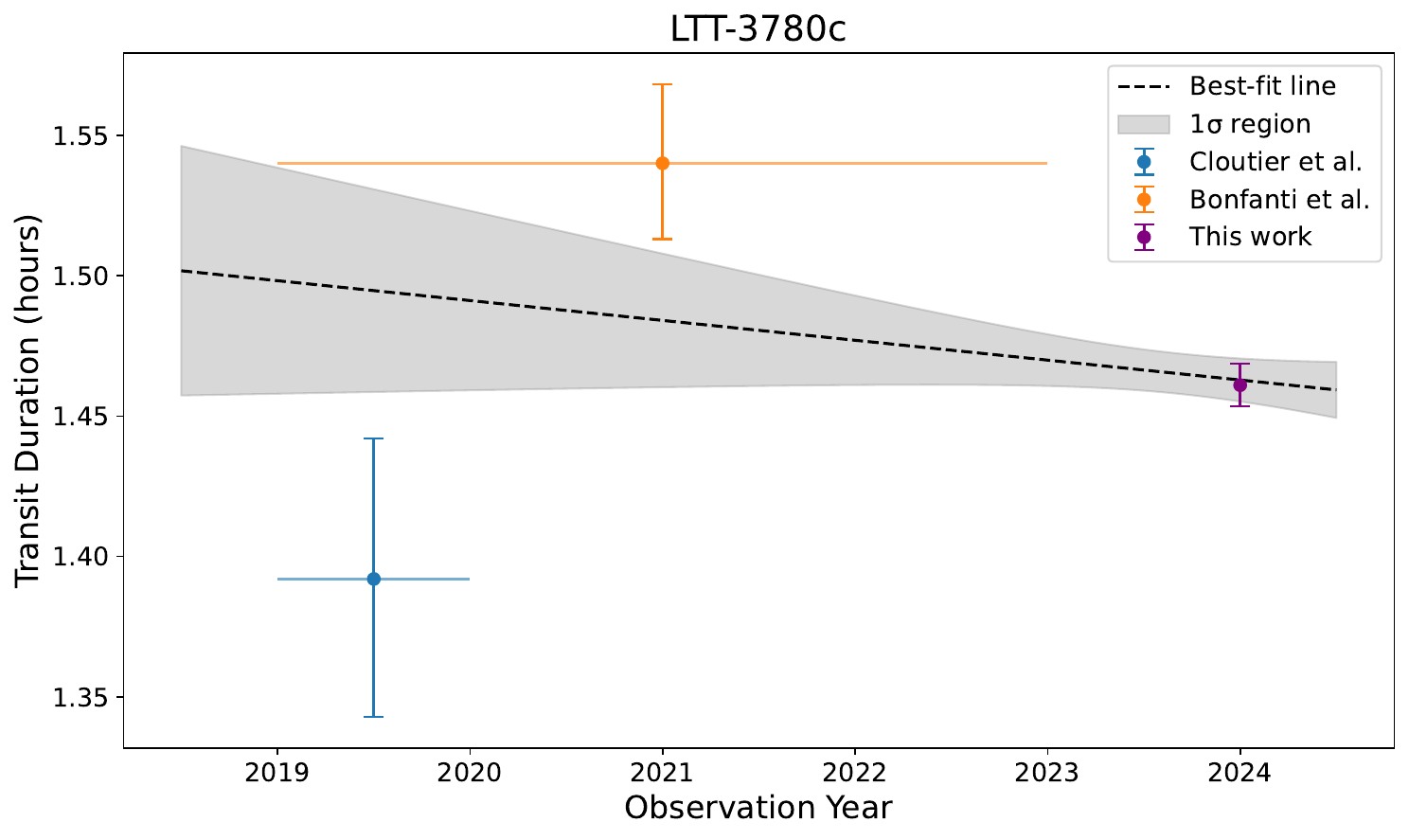}{0.33\textwidth}{}
  \hfill
}

\caption{Time evolution of transit duration for L98-59, GJ-1132, LHS-1140, LP791-18, TOI-270, TOI-178, and LTT 3780.}
\label{fig:TDV_TOI178_LTT}
\end{figure*}

Figures \ref{fig:TDV_TRAPPIST1}, \ref{fig:TDV_GJ_TOI_K2_L9859b}, and \ref{fig:TDV_TOI178_LTT} show the time evolution of transit duration since discovery epoch with the linear best-fit model and 1$\sigma$ region shaded in gray for all 12 M-dwarf multis. 
A flat line corresponds to no change in the transit duration over the observation baseline. We find that no planet exhibits a transit duration variation at the $3\sigma$ level of confidence. The most significant detection is the increase in the transit duration of TRAPPIST-1d by 0.0108$\pm$0.0046 hours, or  39$\pm$17 seconds.

\begin{table*}
\centering
\caption{Constraints on Transit Duration Variation \label{table:planetsampletimeevolution}}

\begin{minipage}[t]{0.48\textwidth}
\centering
\footnotesize
\begin{tabular}{ccc}
\hline
Planet Name & Observation Span & Reference \\
\hline
Trappist-1 b & 2015 & \citep{2016Natur.533..221G} \\ 
\nodata & 2016 & \citep{2017Natur.542..456G} \\
\nodata & 2016--2019 & \citep{Ducrot2020} \\ 
\nodata & 2015--2019 & \citep{2021PSJ.....2....1A} \\ 

Trappist-1 c & 2015 & \citep{2016Natur.533..221G} \\
\nodata & 2016 & \citep{2017Natur.542..456G} \\
\nodata & 2016--2019 & \citep{Ducrot2020} \\
\nodata & 2015--2019 & \citep{2021PSJ.....2....1A} \\

Trappist-1 d & 2016 & \citep{2017Natur.542..456G} \\
\nodata & 2016--2019 & \citep{Ducrot2020} \\
\nodata & 2015--2019 & \citep{2021PSJ.....2....1A} \\

Trappist-1 e & 2016 & \citep{2017Natur.542..456G} \\
\nodata & 2016--2019 & \citep{Ducrot2020} \\
\nodata & 2015--2019 & \citep{2021PSJ.....2....1A} \\

Trappist-1 g & 2016 & \citep{2017Natur.542..456G} \\
\nodata & 2016--2019 & \citep{Ducrot2020} \\
\nodata & 2015--2019 & \citep{2021PSJ.....2....1A} \\

Trappist-1 h & 2016 & \citep{2017Natur.542..456G} \\
\nodata & 2016--2017 & \citep{2017NatAs...1E.129L} \\
\nodata & 2016--2019 & \citep{Ducrot2020} \\
\nodata & 2015--2019 & \citep{2021PSJ.....2....1A} \\

GJ-9827d & 2017 & \citep{2017AJ....154..266N} \\
\nodata & 2017 & \citep{2018AJ....155...72R} \\ 
\nodata & 2017 & \citep{PrietoArranz2018} \\
\nodata & 2017 & \citep{2019MNRAS.484.3731R} \\

TOI-776b & 2019--2020 & \citep{TOI776bDiscovery2021} \\
\nodata & 2019--2021 & \citep{Fridlund2024} \\
\hline
\end{tabular}
\end{minipage}%
\hfill
\begin{minipage}[t]{0.48\textwidth}
\centering
\footnotesize
\begin{tabular}{ccc}
\hline
Planet Name & Observation Span & Reference \\
\hline
TOI-776c & 2019--2020 & \citep{TOI776bDiscovery2021} \\
\nodata & 2019--2021 & \citep{Fridlund2024} \\
GJ-357b & 2019 & \citep{GJ357bdiscovery2019} \\
\nodata & 2019--2022 & \citep{2023AJ....165..134O} \\

K2-18b & 2014 & \citep{2019ApJS..244...11K} \\
\nodata & 2014--2018 & \citep{2017ApJ...834..187B} \\

L98-59b & 2018--2019 & \citep{2019AJ....158...32K} \\
\nodata & 2018--2020 & \citep{Demangeon2021} \\
L98-59c & 2018--2019 & \citep{2019AJ....158...32K} \\
\nodata & 2018--2020 & \citep{Demangeon2021} \\
L98-59d & 2018--2019 & \citep{2019AJ....158...32K} \\
\nodata & 2018--2020 & \citep{Demangeon2021} \\

GJ-1132b & 2014--2015 & \citep{2015ESS.....310201B} \\
\nodata & 2023 & \citep{2024ApJ...973L...8X} \\ 

LHS-1140b & 2014--2016 & \citep{2017Natur.544..333D} \\
\nodata & 2018 & \citep{Lillobox2020}\\
\nodata & 2017--2020 & \citep{2024ApJ...960L...3C} \\ 

LP 791-18c & 2015--2019 & \citep{2019ApJ...883L..16C} \\
\nodata & 2019--2021 & \citep{2023Natur.617..701P} \\

TOI-270b & 2018 & \citep{2019NatAs...3.1099G} \\ 
TOI-270d & 2018 & \citep{2019NatAs...3.1099G} \\ 

TOI-178b & 2018--2020 & \citep{TOI178discovery2021} \\ 
TOI-178d & 2018--2020 & \citep{TOI178discovery2021} \\ 
TOI-178g & 2018--2020 & \citep{TOI178discovery2021} \\ 

LTT-3780c & 2019--2020 & \citep{2020AJ....160....3C} \\
\nodata & 2019--2023 & \citep{Bonfanti2024} \\
\hline
\end{tabular}
\end{minipage}

\end{table*}

\begin{deluxetable*}{cccccc}[ht!]
\tablecaption{Parameters for transit duration change\label{table:parametersfortransitdurationchange}}
\tablehead{
    \colhead{System} & 
    \colhead{Planet Name} & \colhead{\shortstack{Transit Duration (h) \\ (discovery)}} & 
    \colhead{\shortstack{Transit Duration (h) \\ (JWST)}} & 
    \colhead{$\Delta\tau$ [hours]} & \colhead{$\Delta\tau$ [$\sigma_{\Delta\tau}$]}}
\startdata
TRAPPIST-1 & TRAPPIST-1b & $0.602\pm0.0077$ & $0.6026\pm0.0015$ & $-0.0017\pm0.0024$ & $0.70\sigma$ \\
\nodata & TRAPPIST-1c & $0.6963\pm0.0135$ & $0.7050\pm0.0017$ & $0.0010\pm0.0030$ & $0.30\sigma$ \\
\nodata & TRAPPIST-1d & $0.8188\pm0.011$ & $0.8264\pm0.0019$ & $0.0108\pm0.0046$ & $2.35\sigma$ \\
\nodata & TRAPPIST-1e & $0.9535\pm0.012$ & $0.9296\pm0.0024$ & $-0.0065\pm0.0047$ & $1.37\sigma$ \\
\nodata & TRAPPIST-1g & $1.14\pm0.011$ & $1.1498\pm0.0022$ & $0.0096\pm0.0053$ & $1.80\sigma$ \\
\nodata & TRAPPIST-1h & $1.28_{-0.033}^{+0.045}$ & $1.274\pm0.004$ & $0.0055\pm0.0092$ & $0.60\sigma$ \\
GJ-9827 & GJ-9827d & $1.6_{-0.078}^{+0.065}$ & $1.287\pm0.011$ & $0.04\pm0.02$ & $1.78\sigma$ \\
TOI-776 & TOI-776b & $2.41_{-0.1}^{+0.11}$ & $2.3178\pm0.0036$ & $-0.08\pm0.05$ & $1.60\sigma$ \\
\nodata & TOI-776c & $2.99_{-0.13}^{+0.16}$ & $2.9099\pm0.0088$ & $-0.033\pm0.046$ & $0.71\sigma$ \\
GJ-357 & GJ-357b & $1.53_{-0.11}^{+0.12}$ & $1.2964\pm0.0025$ & $-0.18\pm0.10$ & $1.74\sigma$ \\
K2-18 & K2-18b & $2.628^{+0.055}_{-0.067}$ & $2.66\pm0.01$ & $0.000\pm0.031$ & $0.0023\sigma$ \\
L98-59 & L98-59b & $1.02_{-0.13}^{+0.17}$ & $0.9901\pm0.0042$ & $-0.005\pm0.062$ & $0.08\sigma$ \\
\nodata & L98-59c & $1.24_{-0.17}^{+0.12}$ & $1.2345\pm0.0034$ & $-0.065\pm0.088$ & $0.73\sigma$ \\
\nodata & L98-59d & $0.91_{-0.17}^{+0.77}$ & $0.7225\pm0.0077$ & $-0.14\pm0.18$ & $0.77\sigma$ \\
GJ-1132 & GJ-1132b & $0.783\pm0.0233$ & $0.7704\pm0.0039$ & $-0.0030\pm0.021$ & $0.15\sigma$ \\
LHS-1140 & LHS-1140b & $2.069\pm0.1274$ & $2.1107\pm0.0041$ & $0.072\pm0.082$ & $0.87\sigma$ \\
LP 791-18 & LP 791-18c & $1.208_{-0.046}^{+0.056}$ & $1.1711\pm0.0005$ & $0.003\pm0.017$ & $0.17\sigma$ \\
TOI-270 & TOI-270b & $1.387_{-0.034}^{+0.040}$ & $1.4033\pm0.0089$ & $0.0148\pm0.0372$ & $0.40\sigma$ \\
\nodata & TOI-270d & $2.148\pm0.018$ & $2.1246\pm0.0048$ & $-0.023\pm0.019$ & $1.23\sigma$ \\
TOI-178 & TOI-178b & $1.692_{-0.086}^{+0.056}$ & $1.6910\pm0.0172$ & $0.0025\pm0.0723$ & $0.04\sigma$ \\
\nodata & TOI-178d & $2.346_{-0.046}^{+0.047}$ & $2.3644\pm0.0108$ & $0.0165\pm0.0486$ & $0.34\sigma$ \\
\nodata & TOI-178g & $2.167_{-0.082}^{+0.09}$ & $2.3125\pm0.0166$ & $0.146\pm0.085$ & $1.72\sigma$ \\
LTT-3780 & LTT-3780c & $1.392_{-0.049}^{+0.05}$ & $1.4617\pm0.0076$ & $-0.031\pm0.040$ & $0.77\sigma$ \\
\enddata
\label{tbl:transit_duration }
\end{deluxetable*}

{\setlength{\abovecaptionskip}{4pt}      
 \setlength{\dbltextfloatsep}{8pt}       
}

\subsection{Comparison Against Theoretical Predictions} \label{subsec:comparisonbetweentheoreticalprediction}

We consider here, to first order, the expected signal size of transit duration variations in our sample. \cite{guerrero_plausibility_2024} estimated the Laplace-Lagrange frequencies $g$ for pairs of neighboring planets in 280 M dwarf multiplanet systems, employing the  framework of \cite{Millholland19_obliquity}. That work employed mass estimates generated from the mass-radius relationships in \cite{ChenKipping18}. They found $g\sim10^{-2}-10^{-1}$ deg yr$^{-1}$ to be typical, similarly to \cite{Millholland19_obliquity} for multi-planet systems orbiting FGK dwarfs. With this estimated $g$ in hand, it's possible to model the predicted degree of duration variation driven by precession (assuming, in this case, that $g=\dot{\Omega})$. We aim ultimately to identify a distribution of possible $\Delta\tau=\tau_{\textrm{finish}}-\tau_{\textrm{start}}$ for each planet, assuming the $g$ values derived by \cite{guerrero_plausibility_2024}. 

\cite{Millholland21} predicted the TDV signal in \textit{Kepler} multi-planet systems across the 4 year mission baseline, laying out a methology for mapping change in transit duration $\dot{\tau}$ to the change in impact parameter $\dot{b}$. We adopt their methodology and notation conventions here, with the simplification of assuming $\dot{e}=\dot{\omega}=0$ (reasonable for M dwarf multis, per \citealt{sagear_orbital_2023}). It's possible to arrive at an estimate for $\dot{b}$, making a set of assumptions for the fixed angle between the invariable plane and the line of sight $\beta$, the evolving angle between the planet orbital plane and the line of sight $\alpha$, and finally the inclination $i$ between the planet's orbiting plane and the invariable plane (we assume $i$ is fixed per \citealt{Millholland21}). The change in transit duration $\dot{b}$ follows from the evolution of $\dot{\Omega}$ according to

\begin{equation}
\dot{b}=\frac{a}{R_{\star}}\left(\dot{\Omega}\sin{i}\cos{\beta}\sin{\Omega} \right),
\label{eq:bdot}
\end{equation}

where the relationship between these angles is given by

\begin{equation}
\sin{\alpha}=-\sin{i}\cos{\Omega}\cos{\beta}+\cos{i}\sin{\beta}.
\label{eq:sinalpha}
\end{equation}

At any given time, the impact parameter can be calculated from 

\begin{equation}
    b=\frac{a\sin{\alpha}}{R_{\star}},
    \label{eq:b}
\end{equation}

assuming $R_{p}<<R_{\star}$. Again, we note that these expressions have been simplified from their counterparts in \citealt{Millholland21} by assuming that the orbits are strictly circular. In this case, the planet-star separation during transit, $r_{\textrm{mid}}$, is always $a$ (and thus $\dot{r}_{\textrm{mid}}$=0). 

We first identify the full set of geometries that could reproduce the observed duration at discovery. The TDV signal itself is driven solely by the time evolution of $\Omega(t)=\Omega_0+gt$, but will present differently according to the initial geometry. In this sense, $\beta$ and $i$ act as nuisance parameters: we must specify them to establish consistency, though they do not evolve. We construct a grid in $\{\beta,i, \Omega_{0}\}$,  with $\beta\in[-3^{\circ},3^{\circ}]$, $i\in[-3^{\circ},3^{\circ}]$, and $\Omega_{0}\in[0^{\circ},360^{\circ}]$. By evaluating these parameters on a grid, we implicitly assuming a flat uniform distribution for each (we discuss this assumption in Section \ref{sec:discussion}). Using Equations \ref{eq:sinalpha} and \ref{eq:b} above, we calculate $b$ and $\tau$ at each combination. We identify the set of $\{i,\beta,\Omega_{0}\}$ consistent with the transit duration at the discovery epoch. Once the allowed $\{i,\beta, \Omega_{0}\}$ combinations are identified, we propagate each forward in time using $\dot{\Omega}=g$ (where the duration of evolution is set by the time between discovery and the JWST transit measurement) and compute the corresponding $b(t)$ and $\tau(t)$. This procedure effectively marginalizes over the uncertainty in $\beta$, $i$, and the $\Omega_{0}$ at the discovery epoch, while capturing the full range of possible TDV amplitudes arising from precession of the node. In systems with more than one transiting planet, we allow  $\{i,\beta,\Omega_{0}\}$ to vary fully for both, enforcing no relationship in $\Delta i$ between planets. For some $\{i,\beta,\Omega_{0}\}$ consistent with the initial transit duration, the planet ceases to transit during the observation baseline. We discard these solutions. We are left with a set of $\Delta \tau$ which we know (1) are consistent with the discovery duration and (2) have evolved according to $g$.  

\begin{figure*}
\centering
\gridline{
  \fig{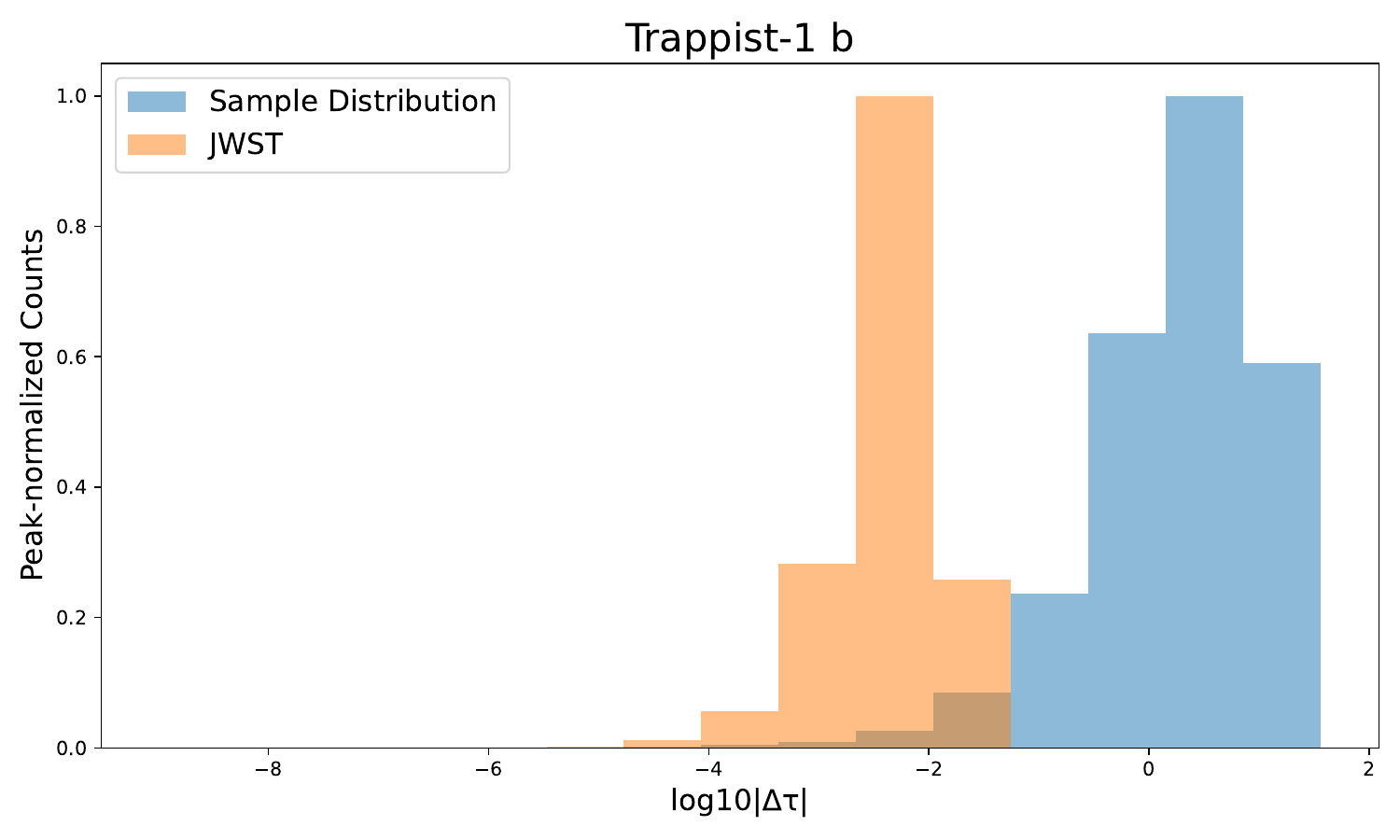}{0.33\textwidth}{\tighttile}
  \fig{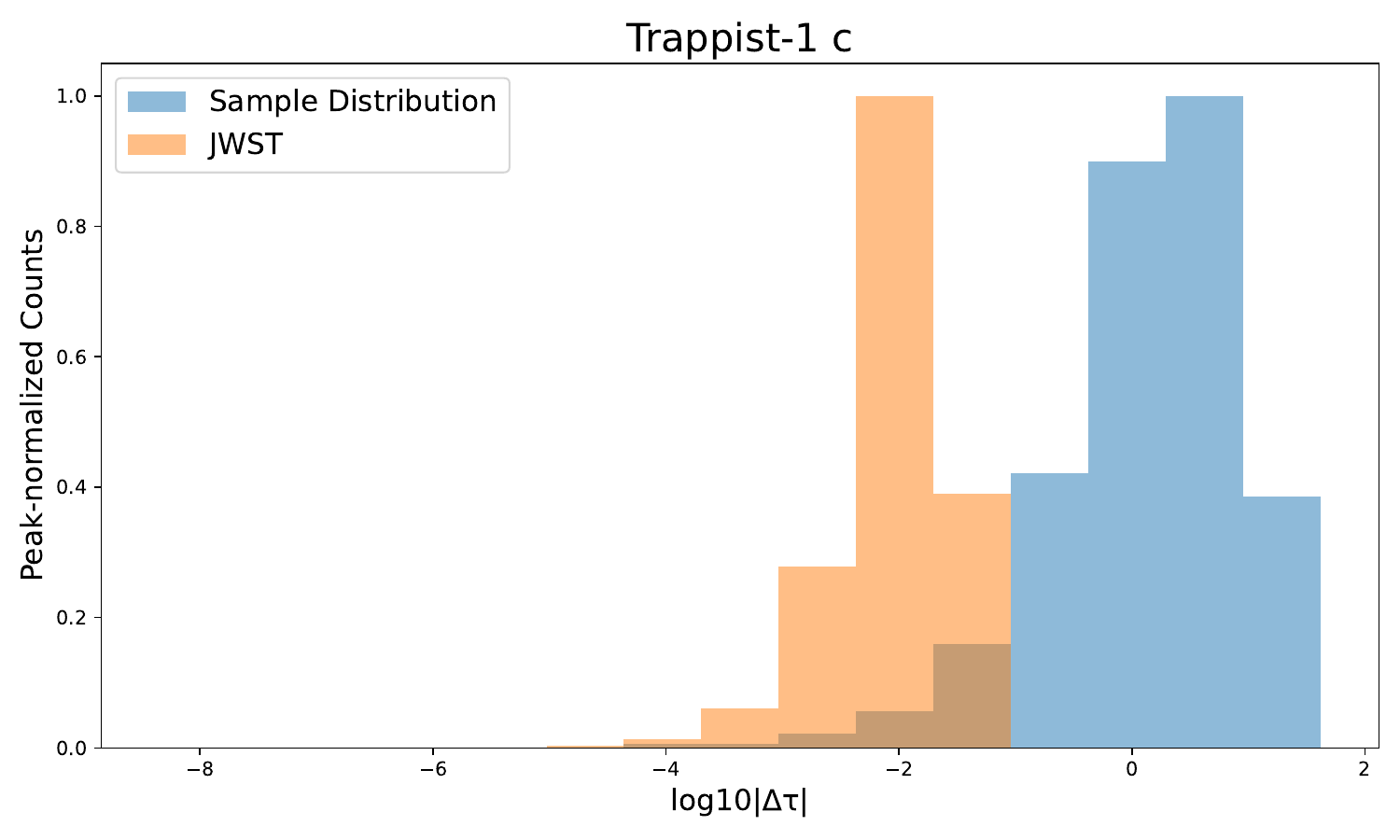}{0.33\textwidth}{\tighttile}
  \fig{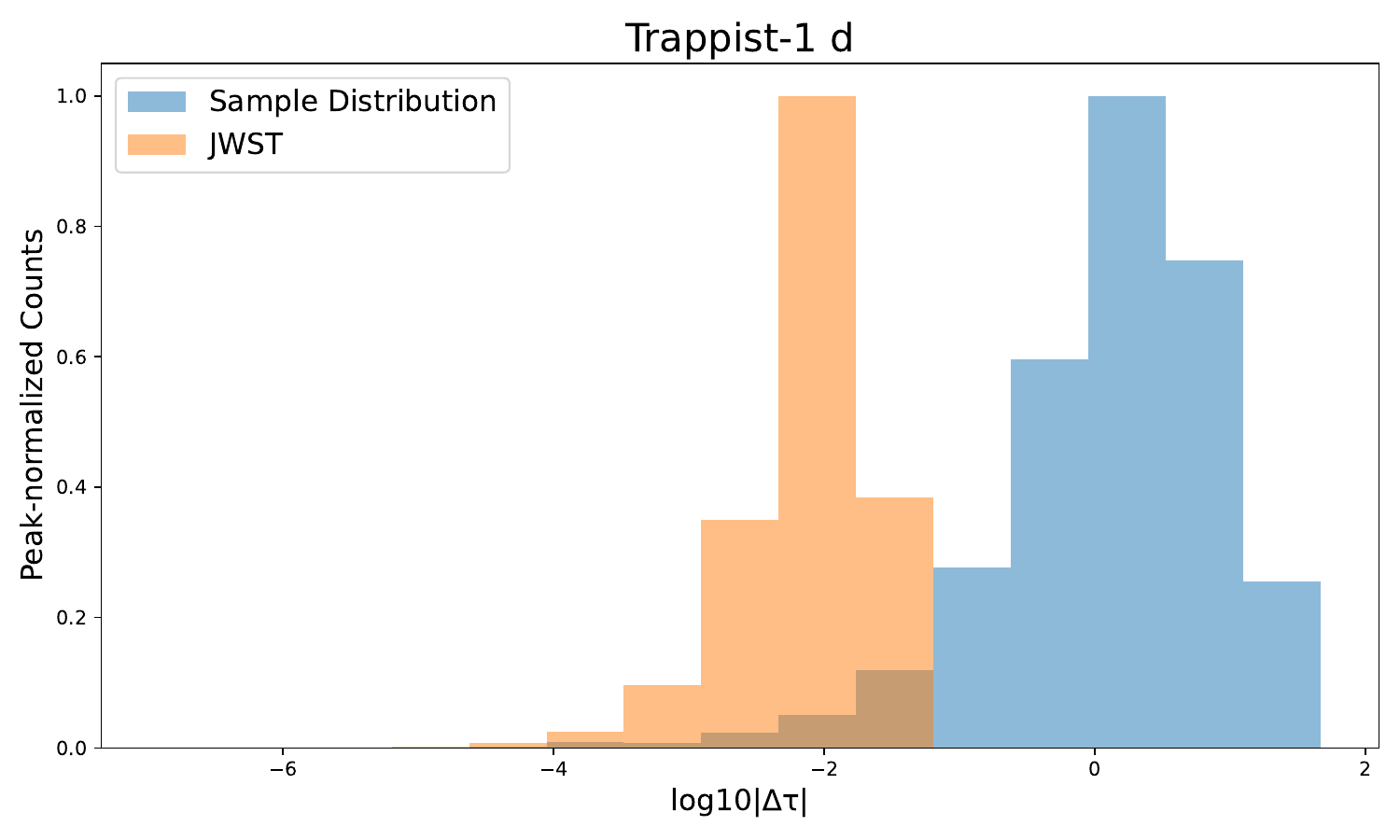}{0.33\textwidth}{\tighttile}
}
\vspace{-30pt}
\gridline{
  \fig{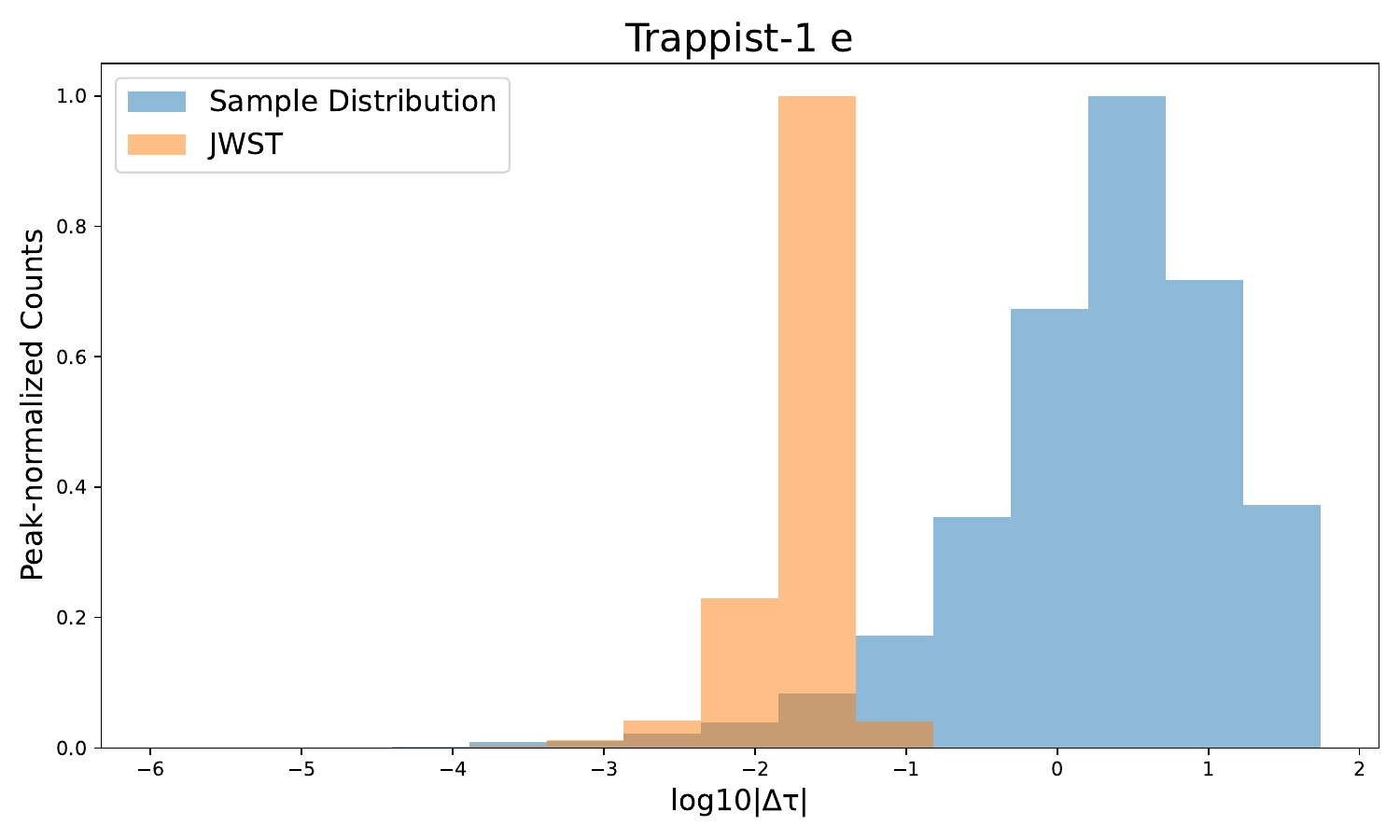}{0.33\textwidth}{\tighttile}
  \fig{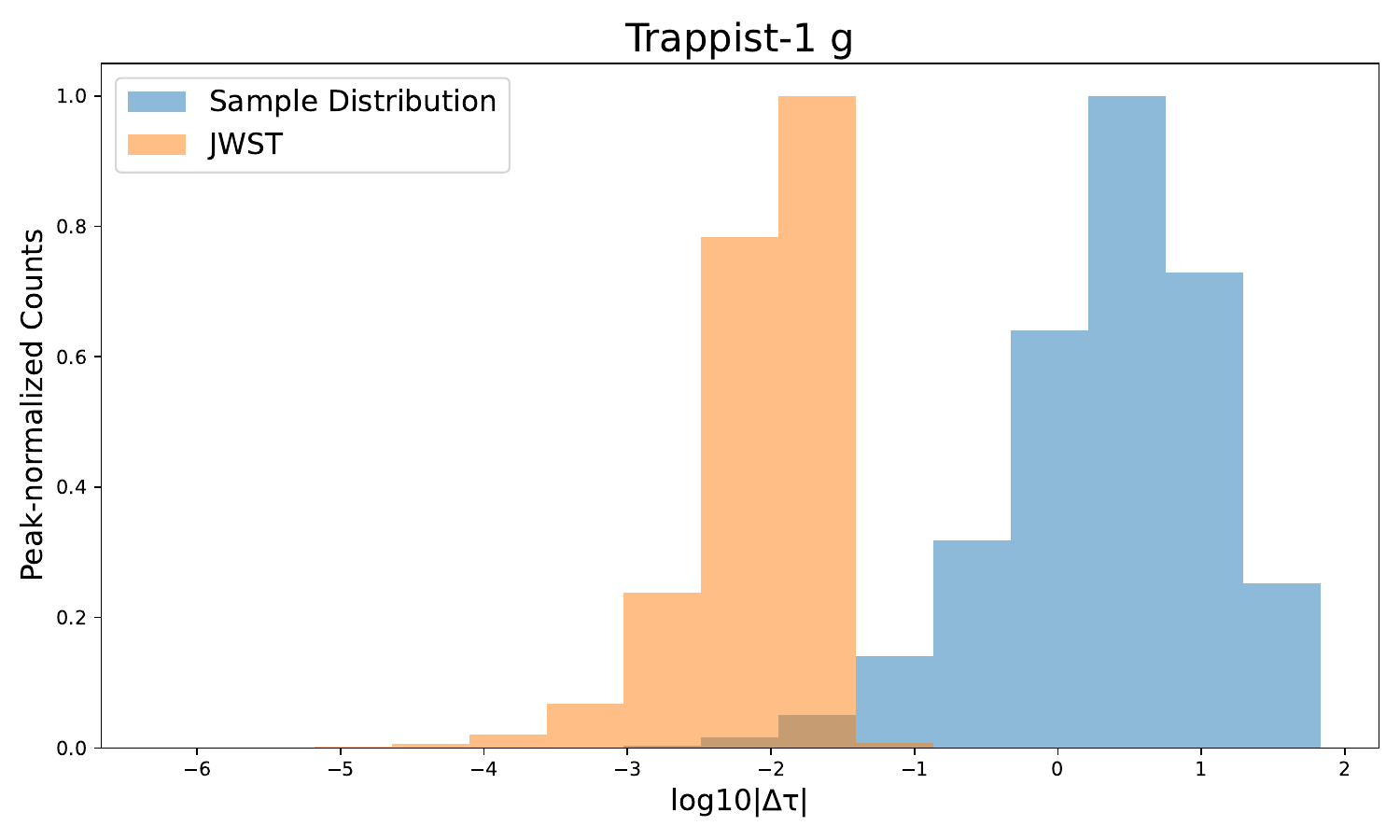}{0.33\textwidth}{\tighttile}
  \fig{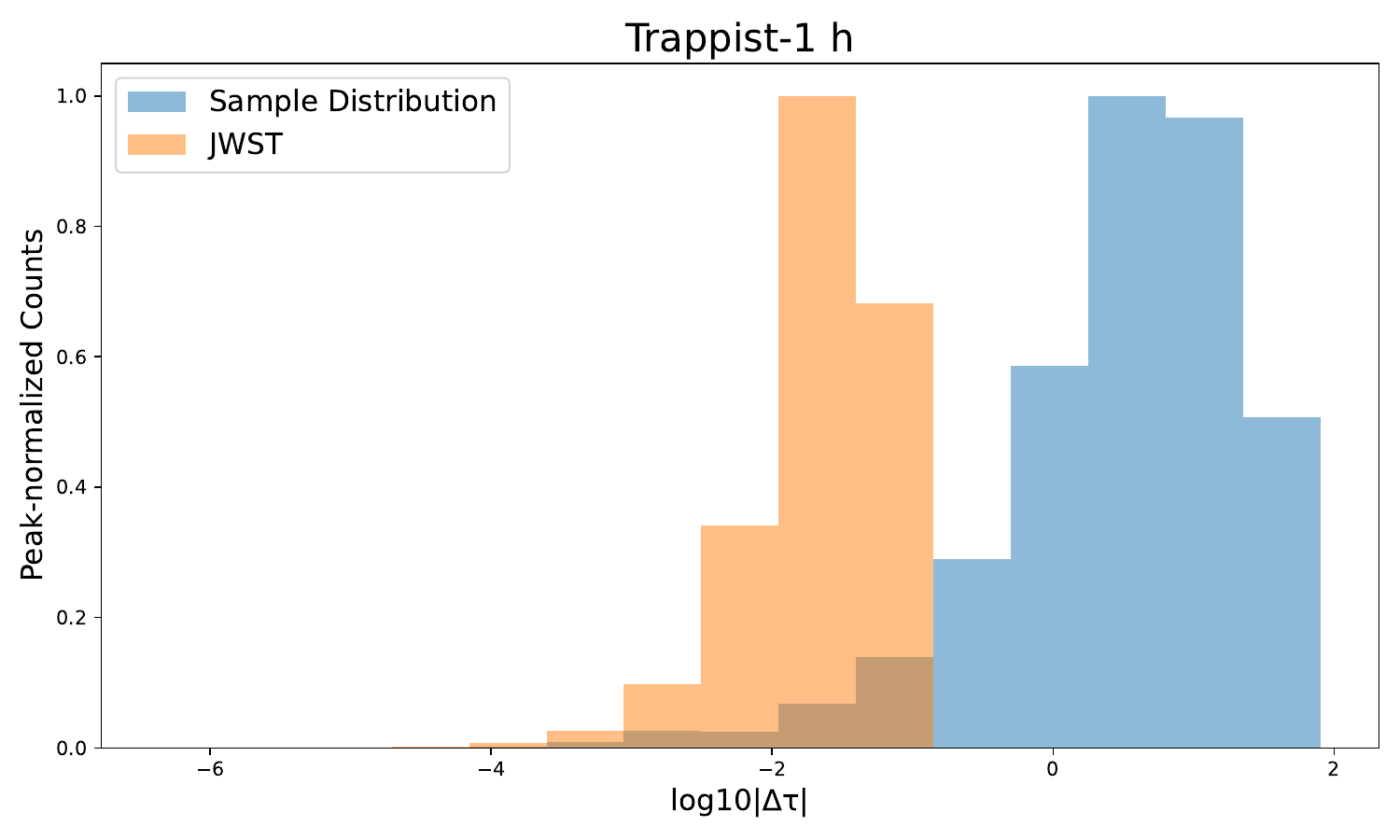}{0.33\textwidth}{\tighttile}
}
\vspace{-30pt}
\gridline{
  \fig{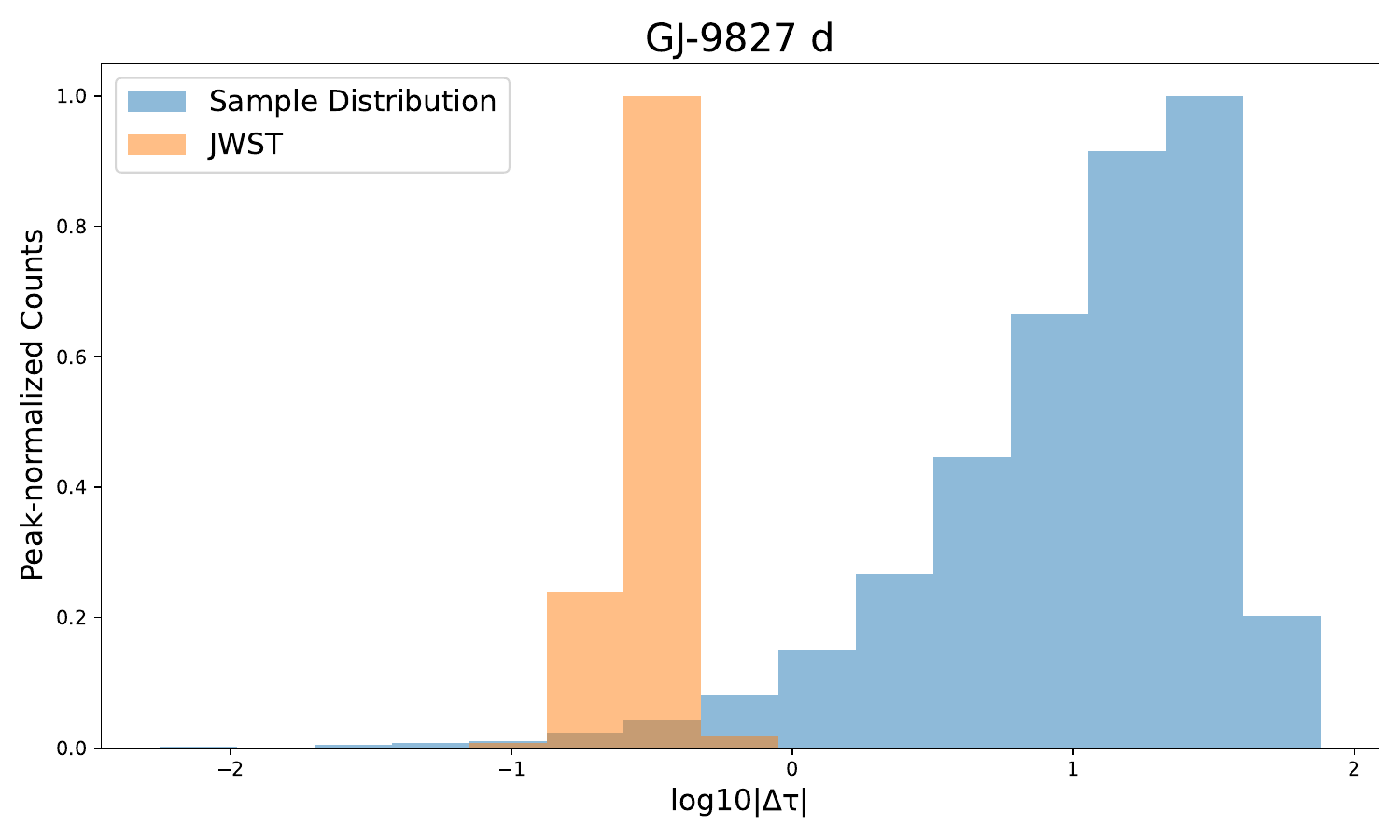}{0.33\textwidth}{\tighttile}
  \fig{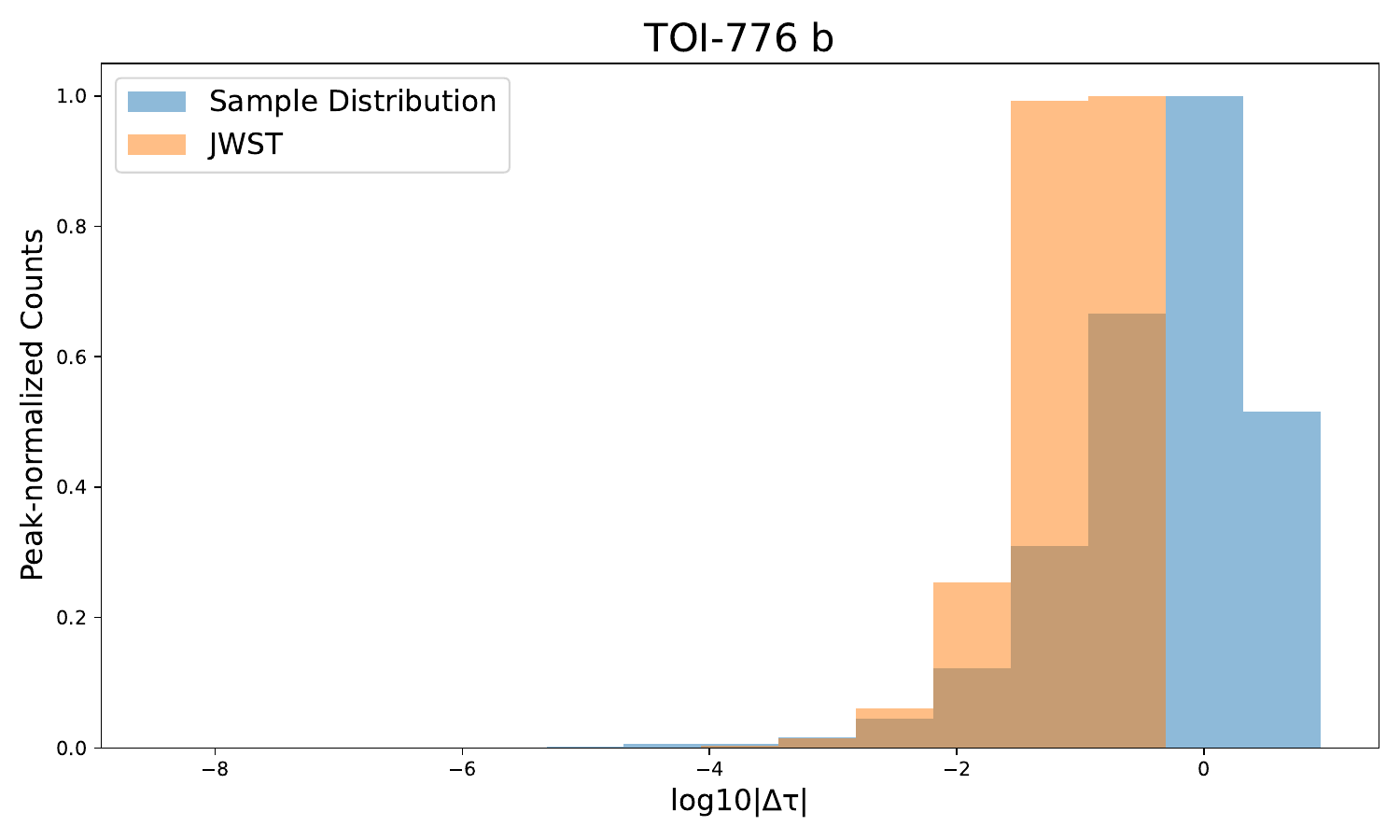}{0.33\textwidth}{\tighttile}
  \fig{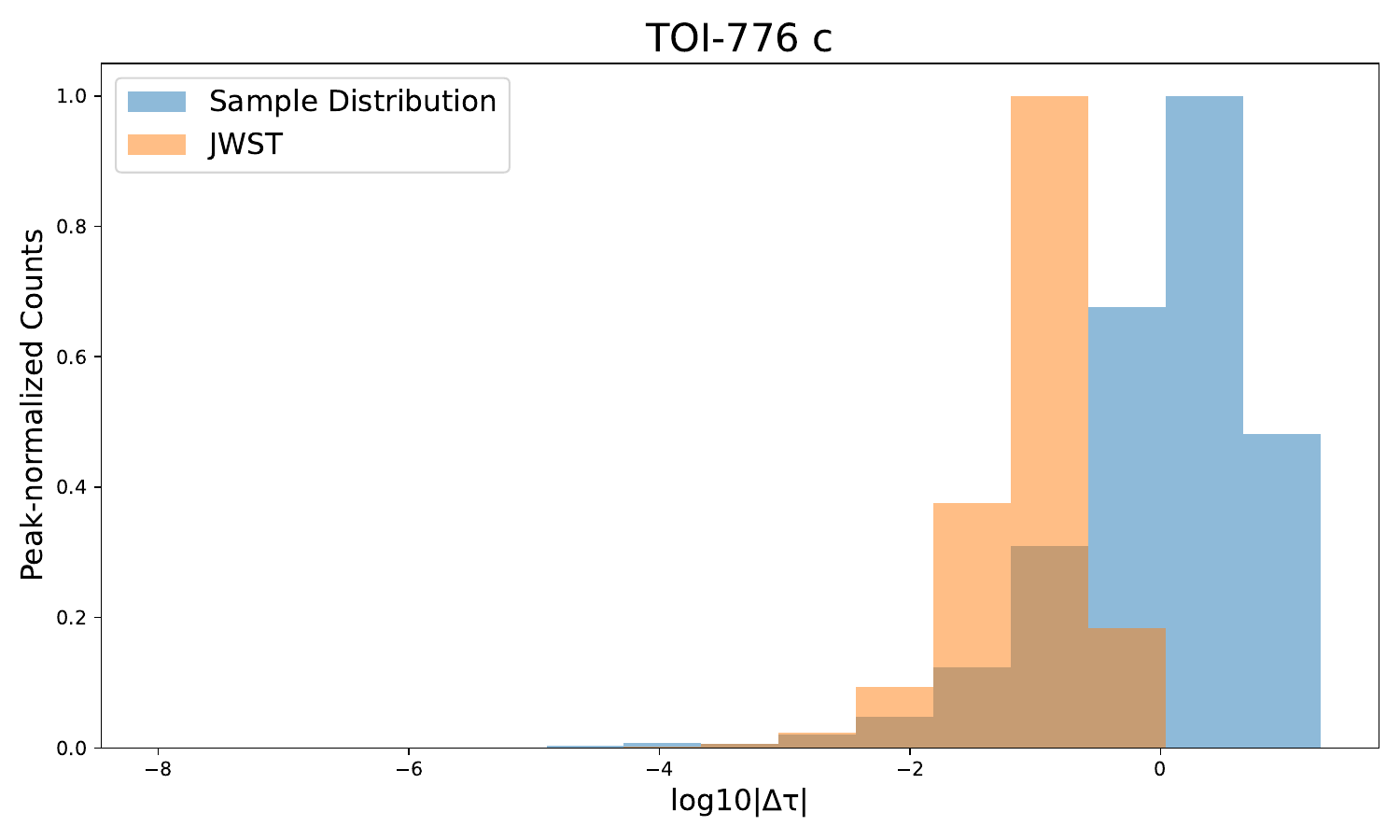}{0.33\textwidth}{\tighttile}
}
\vspace{-30pt}
\gridline{
  \fig{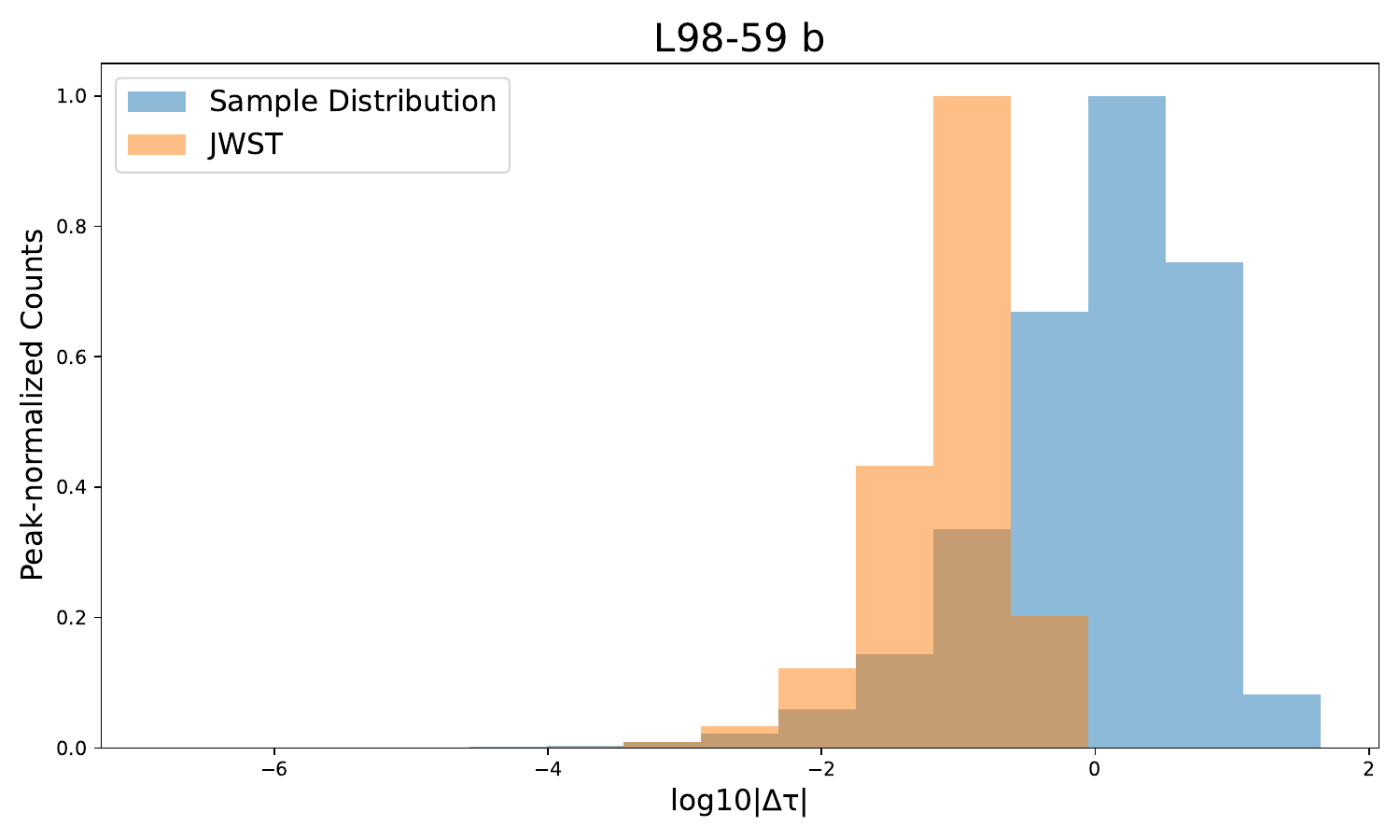}{0.33\textwidth}{\tighttile}
  \fig{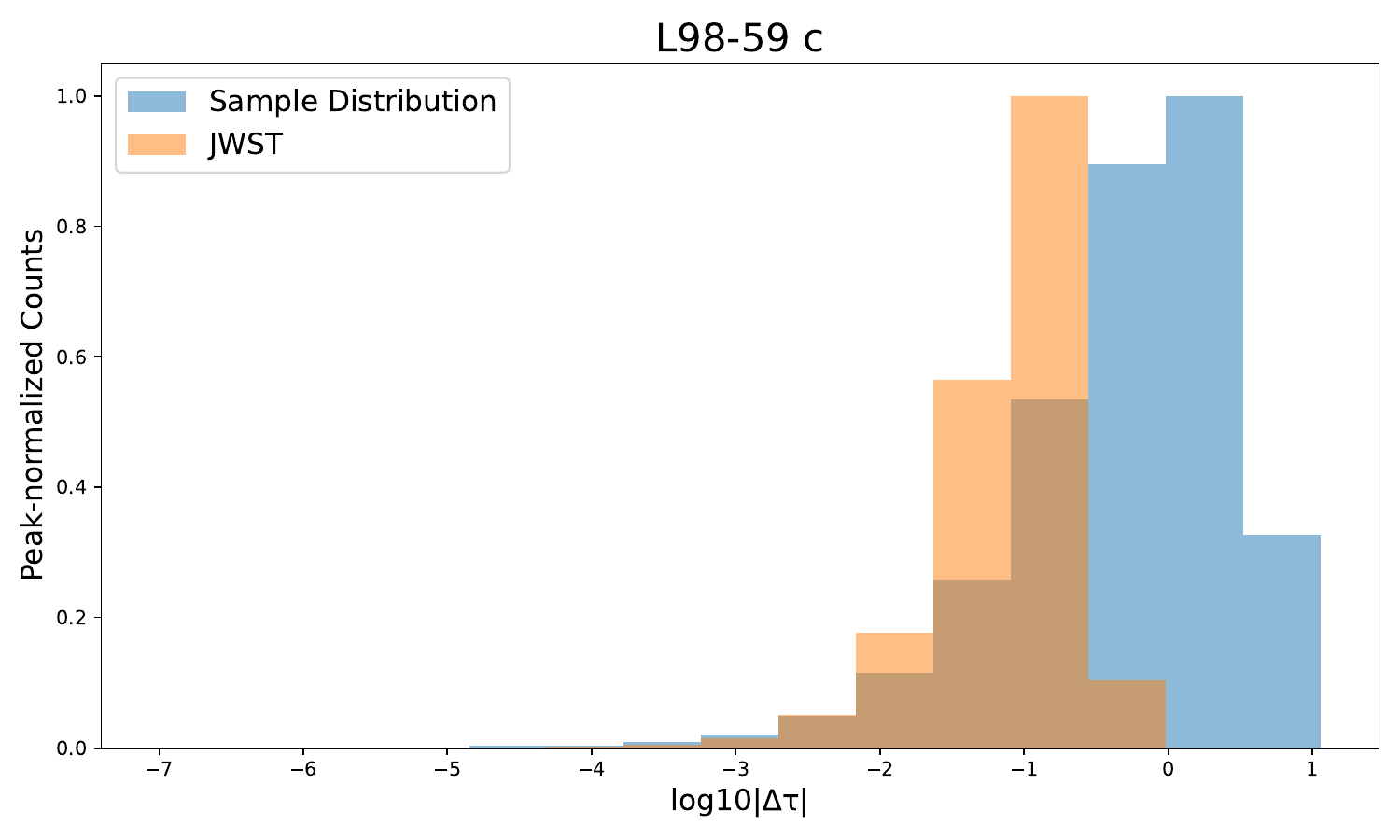}{0.33\textwidth}{\tighttile}
  \fig{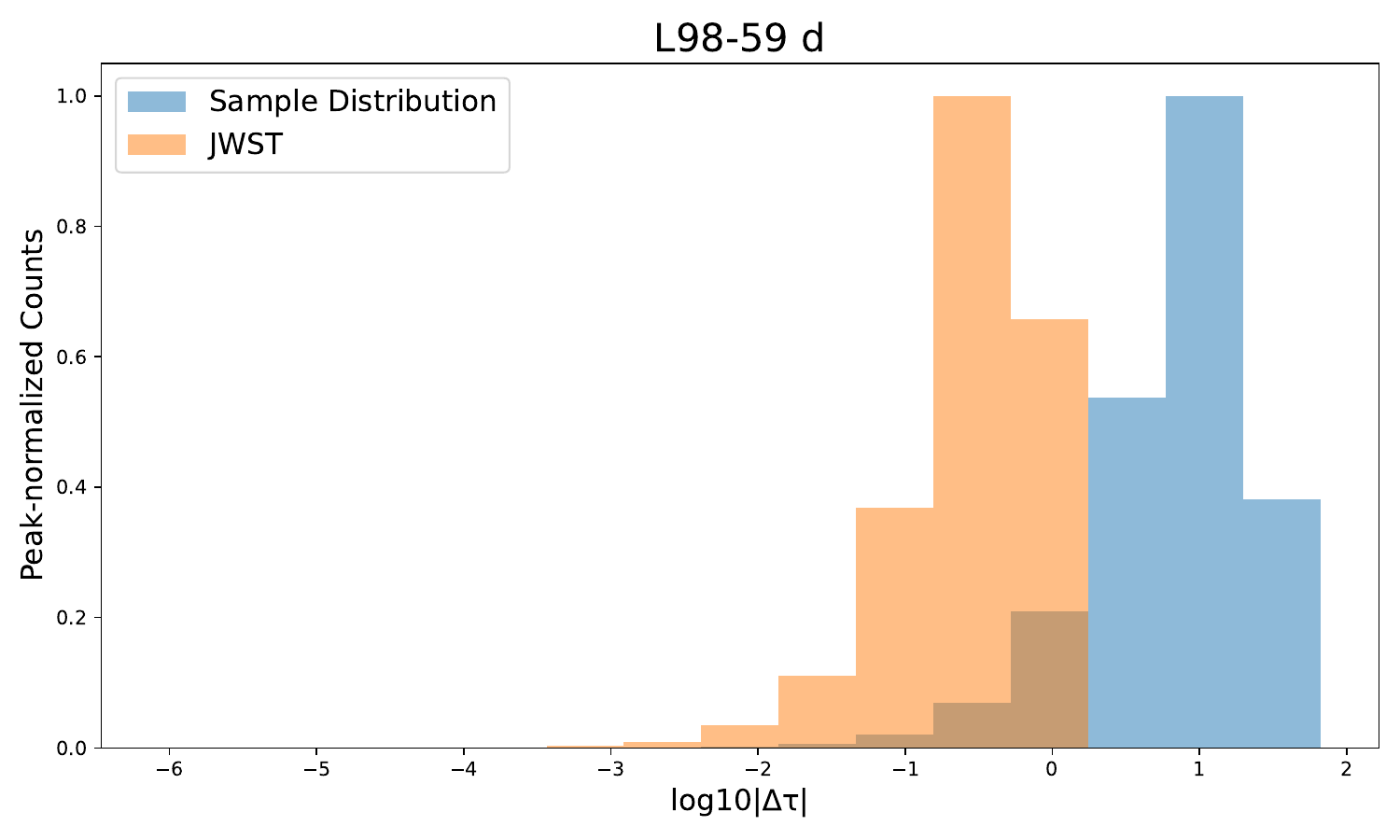}{0.33\textwidth}{\tighttile}
}
\vspace{-30pt}
\gridline{
  \fig{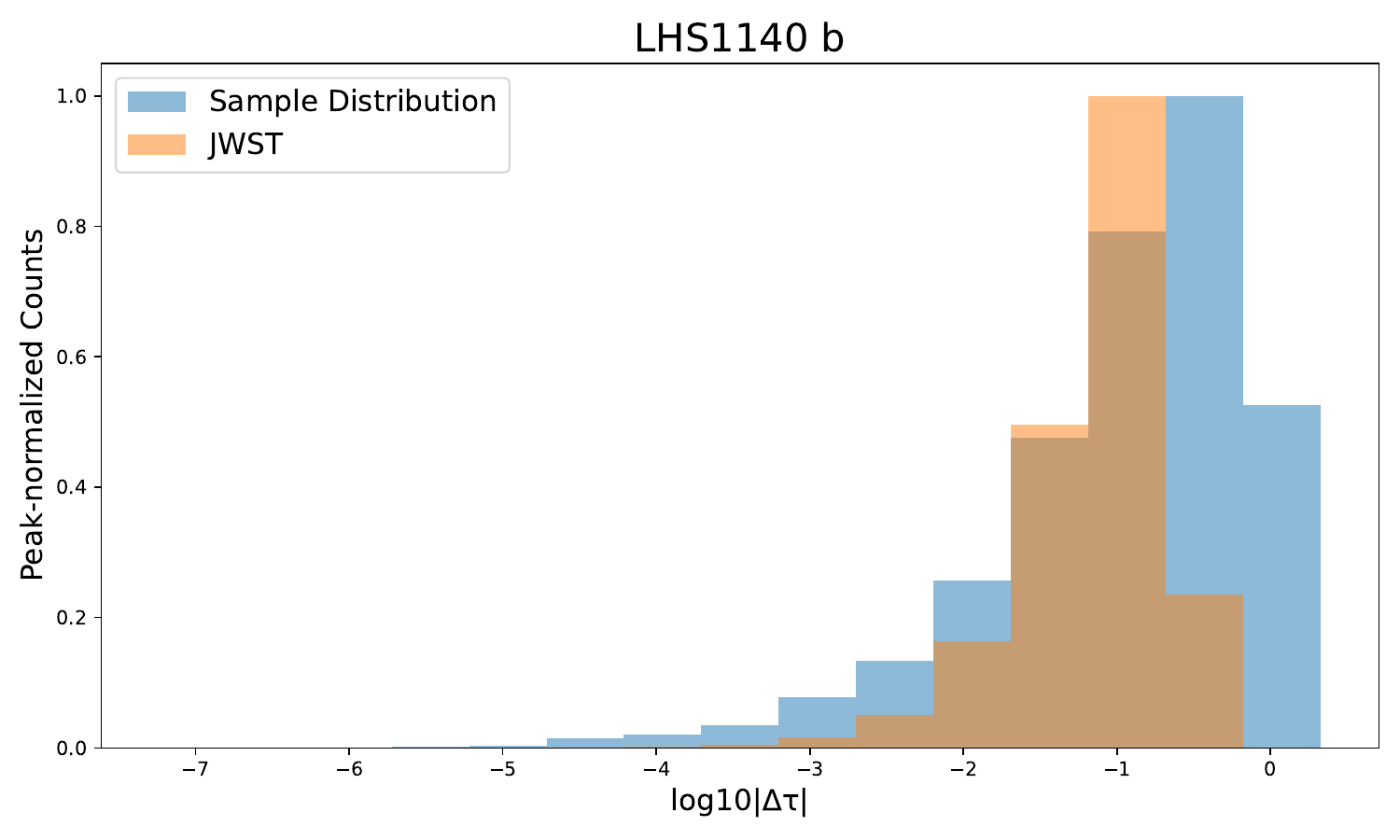}{0.33\textwidth}{\tighttile}
  \fig{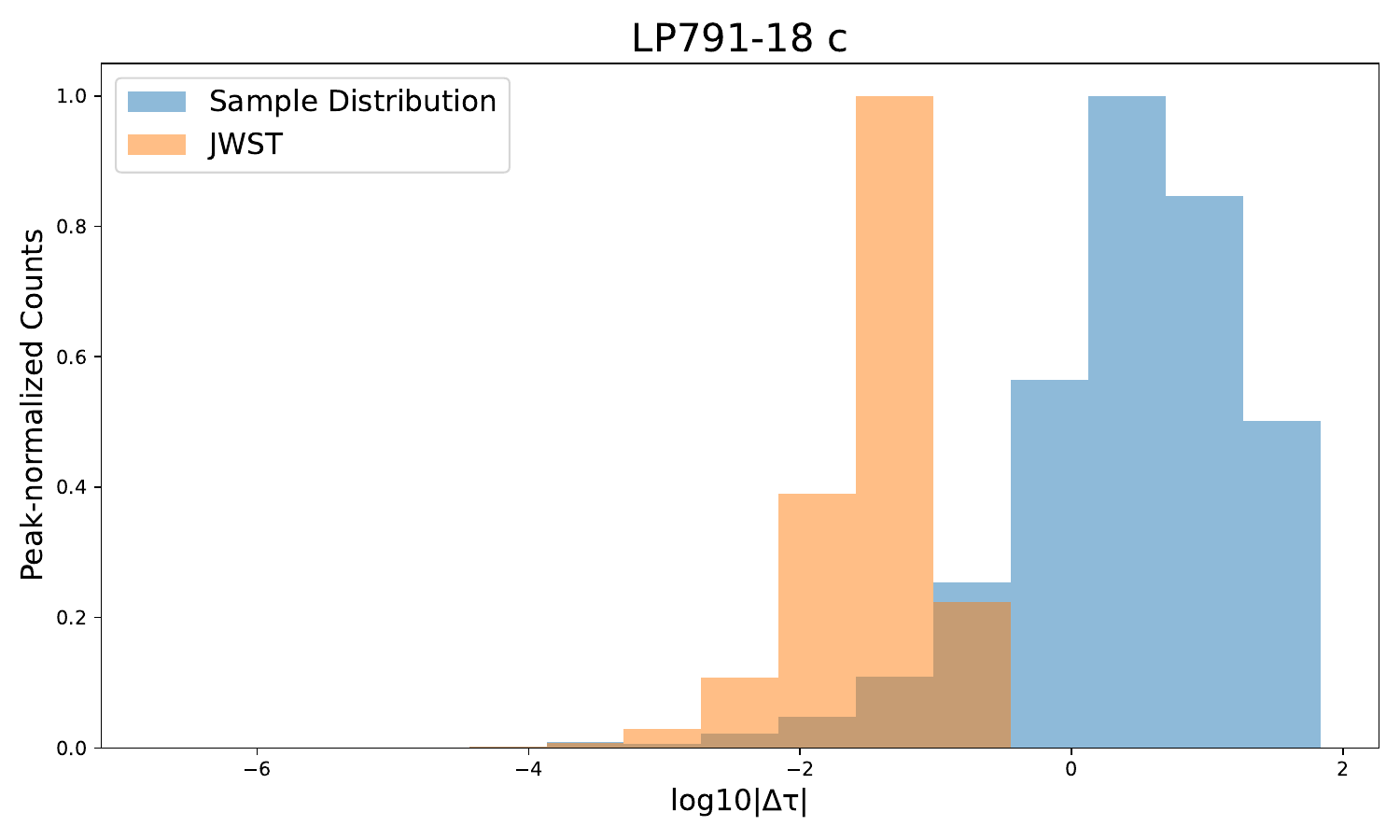}{0.33\textwidth}{\tighttile}
  \fig{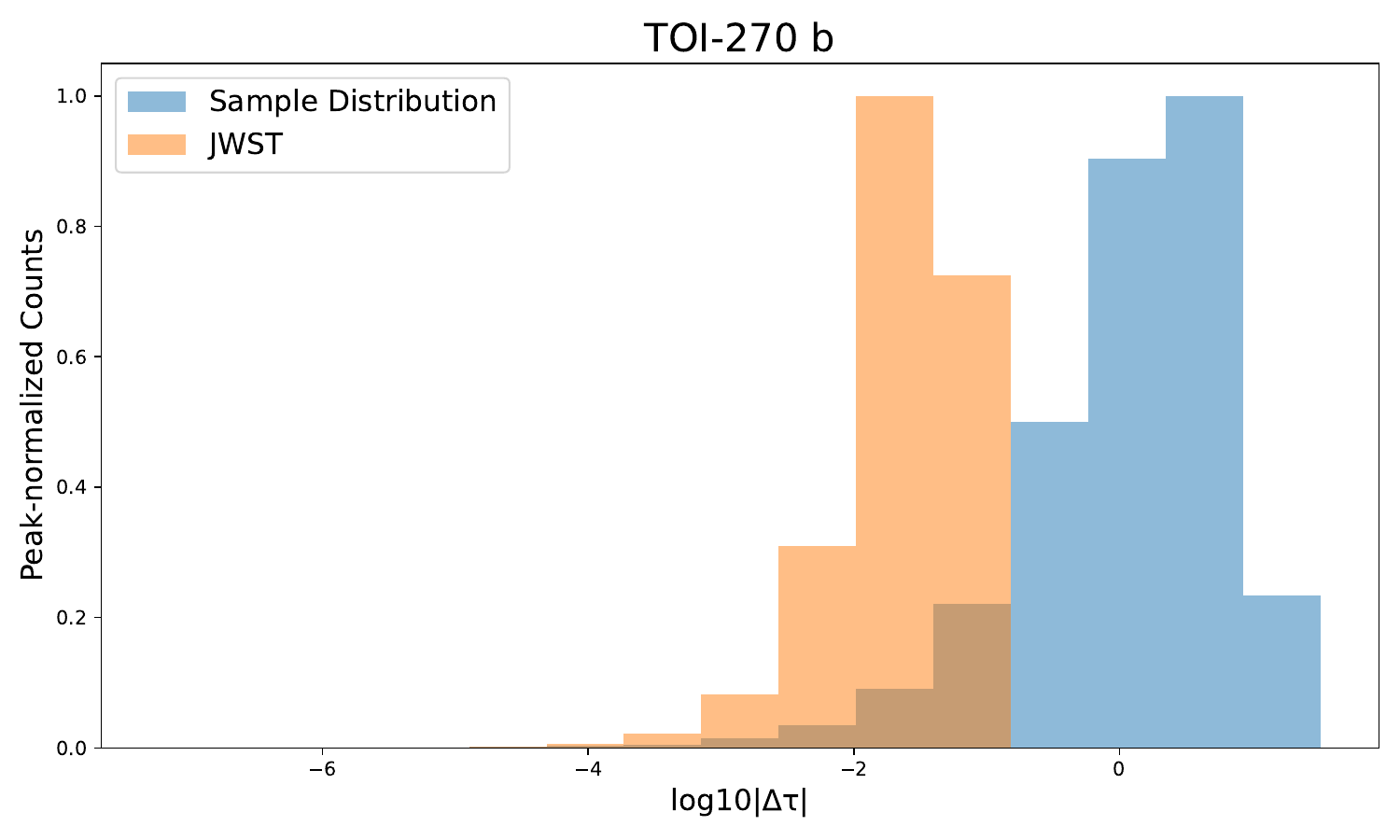}{0.33\textwidth}{\tighttile}
}
\vspace{-30pt}
\gridline{
  \fig{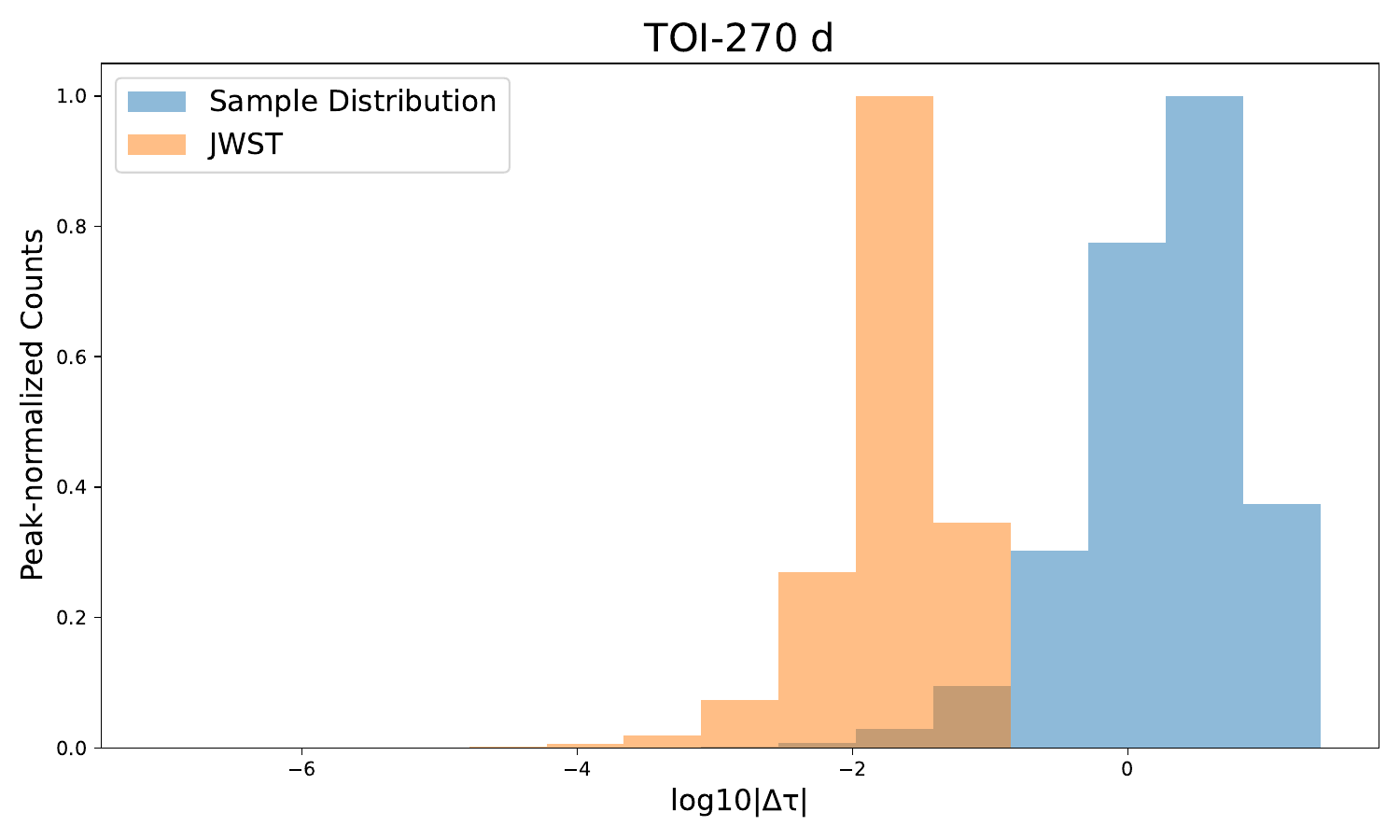}{0.33\textwidth}{\tighttile}
  \fig{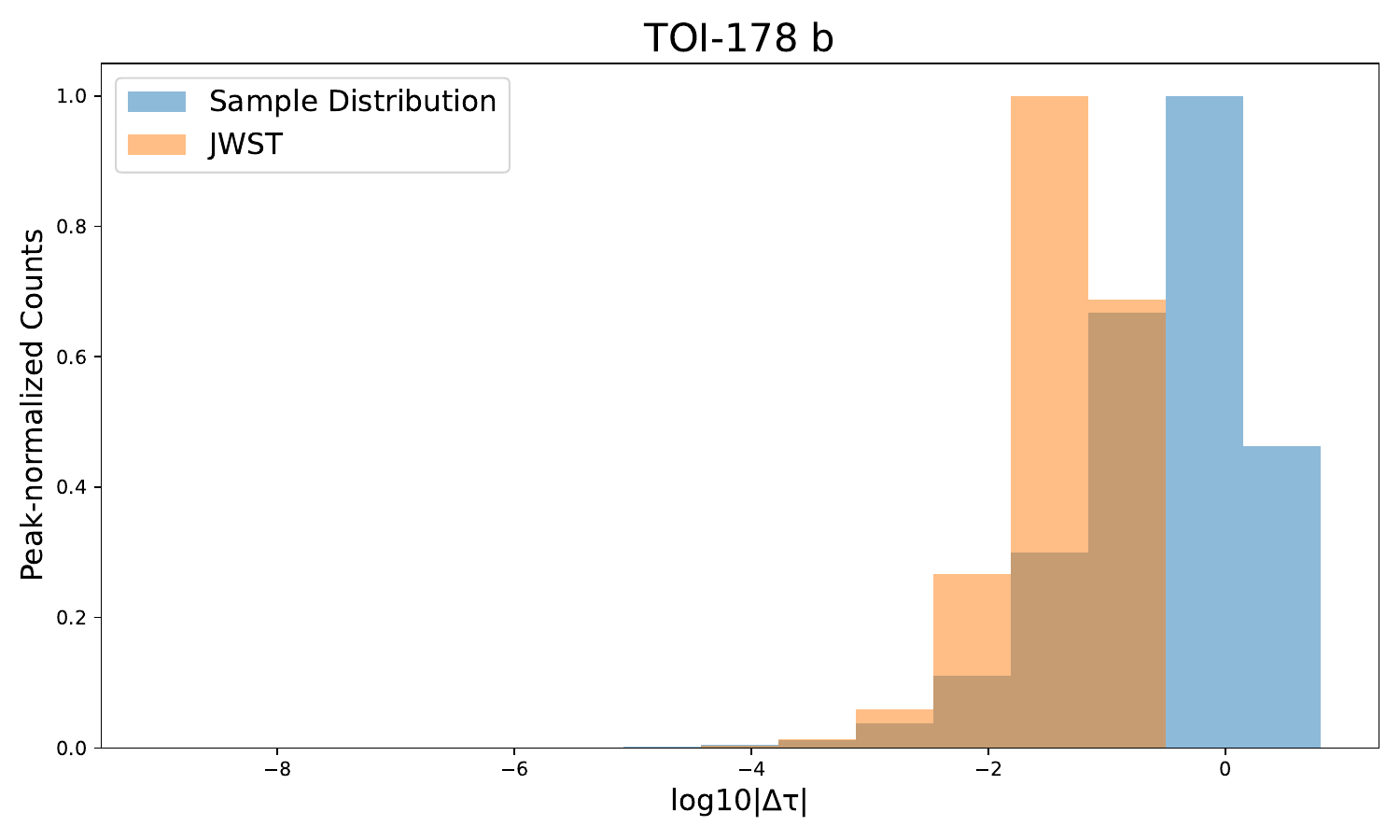}{0.33\textwidth}{\tighttile}
  \fig{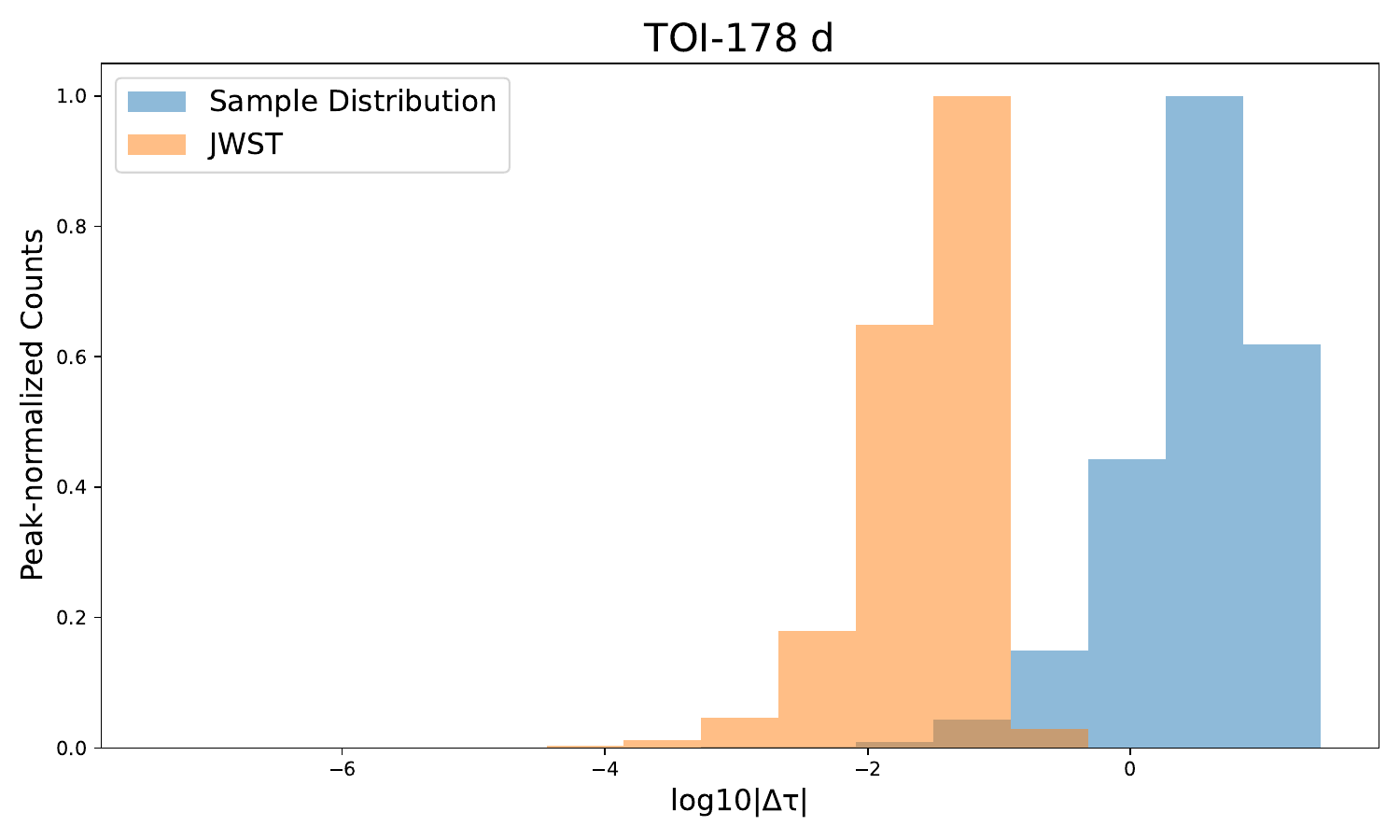}{0.33\textwidth}{\tighttile}
}
\vspace{-30pt}
\gridline{
  \fig{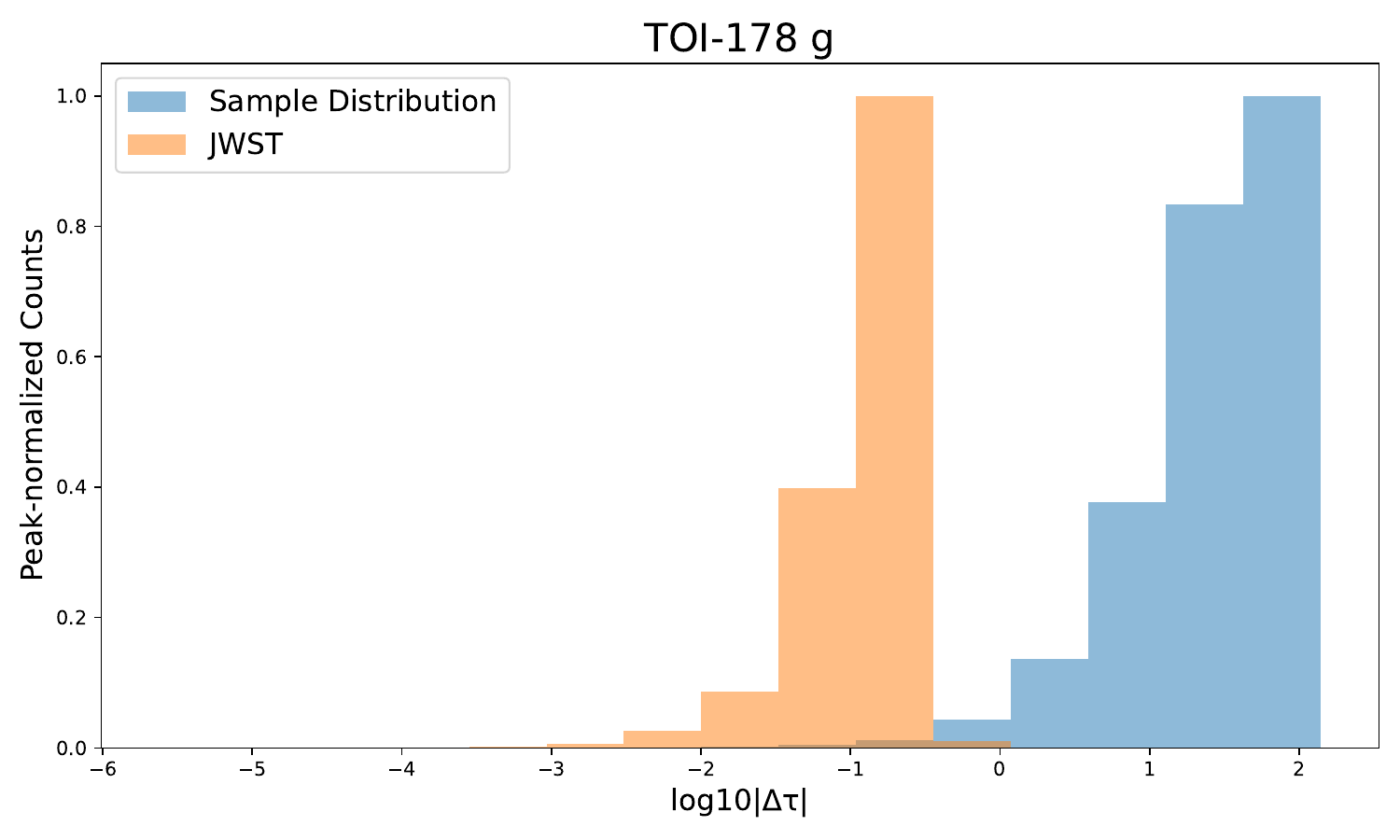}{0.33\textwidth}{\tighttile}
  \fig{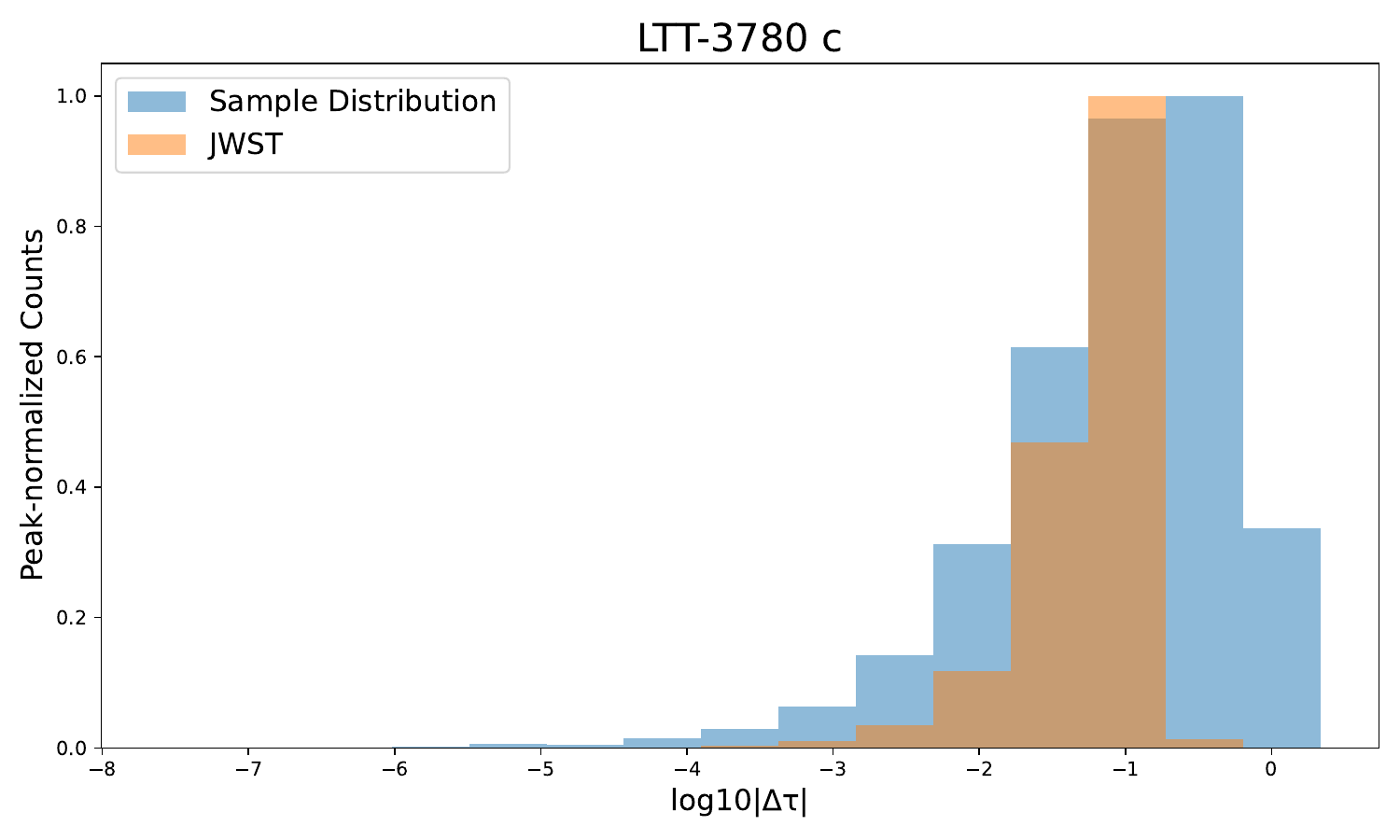}{0.33\textwidth}{\tighttile}
}
\vspace{-30pt}
\caption{For each planet in our sample, the histogram in blue shows the predicted $\Delta\tau$ corresponding to the calculated precession rate $g$ and varying initial geometries. The histogram in orange shows the actual measured $\Delta\tau$ inferred from the analysis describined in Section \ref{subsec:timeevolutionoftransitduration}.}
\label{fig:theoreticalcomparisontrappist1b/c/d}
\end{figure*}

Figure \ref{fig:theoreticalcomparisontrappist1b/c/d} shows the comparison between our simulated TDV outcomes and the actual measured $\Delta \tau$ (given in Table \ref{table:parametersfortransitdurationchange}). We have elected to show $|\Delta \tau|$ so that we can show the dynamic range expected in log space. In all cases, there are at least some possible initial configurations (typically those corresponding to more grazing transit) in which $\Delta \tau$ could have been detectable at the level of $\sim$several-to-ten minutes over the baseline involved. The real observed $\Delta\tau$ distribution reflects our measurement uncertainty per Table \ref{table:parametersfortransitdurationchange}, with the width reflecting the transit duration uncertainties added in quadrature. In a few cases (particularly GJ 9827d, TOI-178d, TOI-178g) the null result is in significant tension with the prediction from theory as calculated above. We discuss these results in Section \ref{sec:discussion}.

{\setlength{\abovecaptionskip}{4pt}      
 \setlength{\dbltextfloatsep}{8pt}       
}

\section{Discussion} \label{sec:discussion}

We discuss here potential interpretations of a null result for the presence of $\sim$min-scale TDVs over a $5-10$ yr baseline in M dwarf multi-planet systems. While we have not detected measureable TDVs in the sample of 23 planets, an argument for the intrinsic rareness of the phenomenon is outside the scope of this study. 

Firstly, our estimation of the Laplace–Lagrange recession rate $g$ relies upon several assumptions, with variations in those assumptions readily pushing the frequency below detectability in this sample. A lower $g$ could naturally arise if the true planetary masses are smaller than the values assumed in our calculation, since $g$ scales approximately with the perturbing planet’s mass; reducing those masses decreases the mutual torque between planets and correspondingly suppresses the precession rate. In addition, our calculation considers only the nodal frequencies associated with immediately adjacent planet pairs. Including modes dominated by more widely separated, non-adjacent planets would generally lower the relevant $g$ values, since the Laplace–Lagrange coupling falls off with increasing period ratio. Assessing whether such weaker, more widely spaced modes contribute appreciably to Cassini-State-2 capture would require a dedicated calculation of the capture probability into CS2 for non-adjacent pairs, and of the proximity of their corresponding  $g$ values to the spin precession frequencies-- a task we defer to future work. 


Second in our consideration is that, even if the estimation of $g$ were robust, there exist orbital configurations that would drive the signal size down or preclude the presentation of TDV altogether (such as the $\beta=0$, $b=0$ case). By simply evaluating $\beta$ and $i$ on a grid, we have by default enforced a uniform flat prior on them. It may be that the configurations most favorable to transit detection in the first place are also those with comparatively low TDV, such that the sample of a priori \textit{transiting} multis is biased in the no-detectable-signal direction. We have not included a probabilistic assessment of that possibility, which would involve careful consideration of the detectability of transits as a function of the impact parameter at time of detection. Indeed, there is an appreciably lower probability of original detection among the grazing, high-$b$ transits (see e.g. \citealt{swift_characterizing_2015}) associated with the largest TDV signals. . 

The estimated secular nodal precession rates imply characteristic timescales of order 
1/$g$, over which the planets’ impact parameters evolve. We have eliminated solutions in which the planet ceases to transit since the discovery epoch (over typically 5 years), but we have not eliminated solutions on the basis of whether a given set of $\{\beta, i\}$ values preserves a multitransiting geometry over a secular cycle. Maintaining multiple planets in a simultaneously transiting configuration over these timescales is not guaranteed—particularly for larger values of $\beta$ and mutual inclination—which would suppress the observable TDV signal even if nodal precession is present. This would require computing the full suite of Laplace–Lagrange eigenfrequencies for each system or directly measuring precession rates from short N-body integrations. Either approach would provide more robust constraints on how much of the parameter space producing large TDVs is also consistent with the observed multiplicity. We defer such system-by-system secular modeling to future work; however, we note that enforcing multitransiting stability would likely prune a substantial fraction of the high-$\beta$, high-inclination configurations that generate the largest TDV amplitudes.

A third related possibility is that we may simply be unlucky in our sampling, even if $g$ is correct. Over these short timescales since discovery, we are investigating a short, approximately linear snapshot of the full sinusoidal TDV curve. In a random time sampling of sinusoids, some snapshots will occur near peak or trough. At these epochs, the gradient in transit duration is smallest. 

Taken together, these considerations show that the absence of detectable TDVs in our sample does not yet imply that nodal precession is dynamically unimportant in these systems. Rather, the interpretation of a null detection depends sensitively on assumptions about planetary masses, secular mode structure, geometric priors, and long-term multitransiting stability, as well as on our limited temporal sampling of the underlying TDV waveform. A more definitive statement will require system-specific secular modeling, either through full Laplace–Lagrange analyses or short N-body integrations, to map out the region of parameter space that simultaneously produces large TDVs and maintains multitransiting geometries, along with longer time baselines to probe beyond the linear segment of the TDV curve. Nonetheless, the present constraints already rule out a substantial fraction of the high-impact-parameter configurations that yield the largest expected signals, and they highlight the promise of TDV monitoring as a route to probing orbital precession and possible obliquity states in compact M-dwarf systems.

\section{Conclusion} \label{sec:conclusion}

We conducted a search for transit duration variations (TDVs) in 23 planets across 12 compact M-dwarf multiplanet systems, combining recent high-precision JWST white-light transit durations with archival measurements spanning 3–10 years.

\begin{itemize}

\item For each planet, we fit a linear trend to transit duration versus time and found no statistically significant TDV detections at the $3\sigma$ level. The strongest candidate, TRAPPIST-1d, exhibits a slope discrepant from zero at $2.2\sigma$.

\item We compared these constraints to TDV amplitudes predicted by secular nodal precession at the Laplace–Lagrange frequencies. Our null detection is consistent with low–impact-parameter geometries, which produce TDVs of only a few seconds per decade, below current sensitivity.

\item Higher–impact-parameter configurations, which generate the largest TDVs, are disfavored by our results. Under the assumption of uniformly distributed viewing geometries, at least half of the \{$\beta$, $i$\} parameter space would produce detectable signals that are not observed.

\item Interpretation of the null result depends on several effects, including uncertainties in planetary masses, contributions from non-adjacent secular modes, geometric selection biases inherent to transiting multi-planet systems, and the requirement that systems remain multitransiting over secular timescales. Each of these can reduce or suppress observable TDVs even when nodal precession is present.

\end{itemize}

Because the TDV signal is approximately sinusoidal over timescales of order $1/g$, longer observational baselines are required to detect or rule out the full range of predicted amplitudes. Future surveys that extend the time baseline or combine JWST with ground-based ultra-precise photometry will be capable of probing a larger fraction of the dynamical parameter space.

\begin{acknowledgments}
We thank Christopher Lam, Quadry Chance, Sheila Sagear, Gregory Gilbert, and Nazar Budaiev for valuable discussions that improved this work. We are especially grateful to Yubo Su for insightful conversations which strengthened the interpretation of our results. This work is based on observations made with the NASA/ESA/CSA James Webb Space Telescope. We acknowledge the JWST Guaranteed Time Observations (GTO) teams and General Observer (GO) programs whose observations made this study possible, and we are grateful to the instrument and pipeline teams for developing and supporting NIRSpec BOTS observations and calibration. This research has made use of the NASA Exoplanet Archive and the Mikulski Archive for Space Telescopes (MAST).
\end{acknowledgments}

\facilities{JWST, Exoplanet Archive}

\software{numpy \citep{harris2020array}, pymc3 \citep{exoplanet:pymc3}, matplotlib \citep{caswell_matplotlibmatplotlib_2024}, astropy \citep{robitaille_astropy_2013, collaboration_astropy_2018, collaboration_astropy_2022}, numpyro \citep{bingham_pyro_2019,phan_composable_2019}, ALDERAAN \citep{gilbert_planets_2025}, jwst \citep{bushouse_jwst_2025}
batman \citep{kreidberg_batman_2015}, pandas \citep{the_pandas_development_team_2024_10957263}, arviz \citep{Kumar2019}, xarray \citep{xarray_v0_8_0}, astroquery \citep{2019AJ....157...98G}, math \citep{van1995python}, corner \citep{2016JOSS....1...24F}, exoplanet \citep{exoplanet:exoplanet}, scipy \citep{2020NatMe..17..261V}
}
\bibliography{references}
\bibliographystyle{aasjournalv7}

\end{document}